\documentclass[prl,aps,onecolumn,superscriptaddress]{revtex4}
\usepackage{graphicx}

\newcommand{\beq}{\begin{equation}}
\newcommand{\eeq}{\end{equation}}
\newcommand{\ba}{\begin{array}}
\newcommand{\ea}{\end{array}}
\newcommand{\bean}{\begin{eqnarray*}}
\newcommand{\eean}{\end{eqnarray}}
\newcommand{\bea}{\begin{eqnarray}}
\newcommand{\eea}{\end{eqnarray}}
\newcommand{\bc}{\begin{center}}
\newcommand{\ec}{\end{center}}
\newcommand{\bt}{\begin{table}}
\newcommand{\et}{\end{table}}

\newcommand{\beqno}{\begin{displaymath}}
\newcommand{\eeqno}{\end{displaymath}}

\newcommand{\been}{\begin{enumerate}}
\newcommand{\een}{\end{enumerate}}

\begin{document}

\title{High-Order-Mode Soliton Structures in Two-Dimensional Lattices with Defocusing Nonlinearity}
\author{P. G.\ Kevrekidis}
\affiliation{Department of Mathematics and Statistics, University of
Massachusetts, Amherst MA 01003-4515}
\author{H.\ Susanto}
\affiliation{Department of Mathematics and Statistics, University of
Massachusetts, Amherst MA 01003-4515}
\author{Z.\ Chen}
\affiliation{Department of Physics and Astronomy, San Francisco State University, 
San Francisco, CA 94132}
\affiliation{TEDA Applied Physics School, Nankai University, Tianjin, China}
\
\begin{abstract}
While fundamental-mode discrete solitons have been demonstrated with 
both self-focusing and defocusing nonlinearity, high-order-mode localized 
states in waveguide lattices have been studied thus far only for the 
self-focusing case.
In this paper, the existence and stability regimes 
of dipole, quadrupole and vortex soliton 
structures in two-dimensional lattices induced with a defocusing nonlinearity 
are examined by the theoretical and numerical analysis of a generic envelope 
nonlinear lattice model. In particular, 
we find that the stability of such high-order-mode solitons is quite 
different from that with self-focusing nonlinearity. As a simple example, a 
dipole (``twisted'') mode soliton which may be 
stable in the focusing case becomes 
unstable in the defocusing regime. Our results
may be relevant to other two-dimensional defocusing periodic nonlinear 
systems such as Bose-Einstein condensates with a positive scattering length 
trapped in optical lattices.
\end{abstract}

\maketitle


\section{Introduction}

Ever since the suggestion of optically induced
lattices in photorefractive media such as Strontium Barium Niobate
(SBN) in \cite{efrem}, and its experimental realization in 
\cite{moti1,neshevol03,martinprl04}, there has been an explosive growth in the
area of nonlinear waves and solitons in periodic lattices.
A stunning array of structures has been predicted and 
experimentally obtained in lattices induced with a self-focusing
nonlinearity, including 
(but not limited to)  discrete dipole \cite{dip},
quadrupole \cite{quad}, necklace \cite{neck} and other
multi-pulse patterns (such as e.g. soliton stripes \cite{multi}), 
discrete vortices 
\cite{vortex}, and 
rotary solitons \cite{rings}.
Such structures have a potential to be used as carriers and conduits for
data transmission and processing, in the context of all-optical 
schemes. A recent review of this direction can be found in \cite{moti3}
(see also \cite{zc4}).

Many of these studies in induced lattices were
also triggered by the pioneering work done in fabricated AlGaAs 
waveguide arrays \cite{7}. In the latter setting a multiplicity
of phenomena such as discrete diffraction, Peierls barriers,
diffraction management \cite{7a} and gap solitons \cite{7b} among
others \cite{eis3} were experimentally obtained. These phenomena, in turn,
triggered a tremendous increase also on the theoretical side of
the number of studies addressing such effectively discrete media;
see e.g. \cite{review_opt,general_review} for a number of relevant
reviews.

Finally, yet another area where such considerations and structures
are relevant is that of soft-condensed matter physics, where droplets
of Bose-Einstein condensates (BECs) may be trapped in an (egg-carton) 
two-dimensional optical lattice potential \cite{bloch}. The latter
field has also experienced a huge growth over the past few years,
including the prediction and manifestation of modulational instabilities
\cite{pgk}, the observation of gap solitons \cite{markus} and Landau-Zener
tunneling \cite{arimondo} among many other salient features; reviews of
the theoretical and experimental findings in this area have also been
recently appeared in \cite{konotop,markus2}.

In light of all the above activity, it is interesting to note that
the only structure that has been experimentally observed in 
two-dimensional (2d) lattices in ``defocusing'' media 
consists of self-trapped ``bright'' wave packets (so-called ``staggered''
or gap solitons) excited in the vicinity of the edge of the first
Brillouin zone
\cite{moti1}. 
However more complex 
coherent structures have not yet been explored in lattices with
defocusing nonlinearity and their stability properties
have not yet been examined, to
the best of our knowledge. It
should be mentioned that the defocusing context is accessible 
in the aforementioned settings. E.g., in the photorefractive
lattices, this can be done by appropriate reversal of the applied voltage
to the relevant crystal, 
while in 
BECs, the defocusing nonlinearity corresponds to the most typical
case arising in dilute gases of $^{87}$Rb or $^{23}$Na.

It is the aim of the present work to examine the non-fundamental soliton 
structures (e.g., dipoles, multipoles, and vortices) in lattices with a 
defocusing nonlinearity, and to illustrate the similarities and differences 
in comparison to their counterparts in the focusing case.  In particular, 
we study dipole structures (consisting of two peaks) and quadrupole 
structures (featuring four peaks), as well as vortices of topological 
charge $S=1$ (cf. \cite{vortex}) in a 2D induced lattice with a defocusing 
nonlinearity. These structures will be analyzed in detail for both cases, 
namely, the ``on-site'' excitation (where the center of the structure is 
on an empty lattice site between the excited ones) and 
the ``inter-site'' excitation (where their 
center is between two lattice sites and no empty lattice site
exists between the excited ones).

Our study of these structures will be conducted
analytically and numerically (in the next two sections) in
the context of the most prototypical generic envelope lattice
model, the so-called discrete nonlinear Schr{\"o}dinger
(DNLS)
equation with a defocusing nonlinearity 
\cite{dnls} which is related to all of the above 
contexts \cite{review_opt,konotop}. 
When we find the relevant structures to be unstable, we will
also briefly address the dynamical evolution of the instability,
through appropriately crafted numerical experiments.
Finally, in the last section, we will summarize
our findings and present our conclusions, and the interesting
experimental manifestations that they suggest.

\section{Model and Theoretical Setup}

As our generic envelope model
encompassing the main features of discrete diffraction and 
defocusing nonlinearity we use the two-dimensional (2D) DNLS
equation:
\begin{equation}
i\dot{u}_{\mathbf{n}}=-C \left( \Delta _{2}u\right)
_{\mathbf{n}} + |u_{\mathbf{n}}|^{2}u_{\mathbf{n}},  
\label{DNLS}
\end{equation}
where $u_{\mathbf{n}}$ is a complex amplitude of the electromagnetic wave in
nonlinear optics
\cite{review_opt}, or the BEC wave function at the nodes of
a deep 2D optical lattice \cite{konotop};  $\mathbf{n}$
is the (two-dimensional in the present study)
vector lattice index, and $\Delta_{2} $ the
standard discrete Laplacian. Furthermore, $C $ is the constant
of the intersite coupling (associated with the 
interwell ``tunnelling rate'' \cite{konotop}),
and the overdot stands for the
derivative with respect to the evolution variable, which can be $z$ in optical
waveguide arrays, or the time 
$t$ in the BEC model. We focus on  standing-wave solutions
of the form $u_{\mathbf{n}}=\exp (-i\Lambda t)\phi _{\mathbf{n}}$, with
$\phi _{\mathbf{n}}$ satisfying the 
equation,
\begin{equation}
f(\phi _{\mathbf{n}},C) \equiv -\Lambda \phi _{\mathbf{n}}-C \Delta
_{2}\phi _{\mathbf{n}}+|\phi _{\mathbf{n}}|^{2}\phi _{\mathbf{n}}=0.
\label{steady}
\end{equation}
Perturbing around the solutions of Eq. (\ref{steady})  gives rise
to the linearization operator
\begin{eqnarray}
\mathcal{H}_{\mathbf{n}}^{(C)} = \left(
\begin{array}{cc}
-\Lambda +2|\phi _{\mathbf{n}}|^{2} & \phi _{\mathbf{n}}^{2} \\
\bar{\phi}_{\mathbf{n}}^{2} & -\Lambda +2|\phi _{\mathbf{n}}|^{2}\end{array}\right)
-C \Delta _{2}\left(
\begin{array}{cc}
1 & 0 \\
0 & 1\end{array}\right),  
\label{oper}
\end{eqnarray}
with the overbar denoting complex conjugation. Through an appropriate rescaling
of the equation, we can fix $\Lambda \equiv 1$. Our analysis uses as
a starting point the so-called anti-continuum limit, i.e., the case of 
$C=0$, where 
for the uncoupled sites, 
\begin{eqnarray}
\phi_{\mathbf{n}}=r_{\mathbf{n}} e^{i\theta _{\mathbf{n}}},
\label{AC}
\end{eqnarray}
with the amplitude $r_{\mathbf{n}}$ 
being $0$ or $\sqrt{\Lambda}$, and the phase $\theta _{\mathbf{n}}$ being
an arbitrary constant. Continuation of such a solution to 
nontrivial couplings necessitates that a certain, so-called 
Lyapunov-Schmidt condition be satisfied \cite{dep}. The latter
imposes for  the projection of
eigenvectors of the kernel of $\mathcal{H}_{\mathbf{n}}^{(0)}$
onto the system of stationary
equations to be vanishing. This solvability condition provides
a nontrivial constraint at every ``excited'' (i.e., $r_{\mathbf{n}}\neq0$)
site of the AC limit, namely:
\begin{equation}
-2ig_{\mathbf{n}}(\theta ,C ) \equiv -C e^{-i\theta
_{\mathbf{n}}}\Delta _{2}\phi _{\mathbf{n}}+C e^{i\theta
_{\mathbf{n}}} \Delta _{2}\bar{\phi}_{\mathbf{n}}=0. 
\label{inter}
\end{equation}
It is interesting (and crucial for stability purposes) to note
that this equation has an extra $(-)$ sign in comparison to its
focusing counterpart. 
The derivation of these solvability conditions is especially important
because the corresponding Jacobian 
\begin{eqnarray}
\mathcal{M}_{ij}=
\partial g_{i}/\partial \theta_{j}
\label{jacobian}
\end{eqnarray}
 has eigenvalues $\gamma$ that are 
directly related to the ``regular'' eigenvalues of the linearization problem
$\lambda $, through the equation 
\begin{eqnarray}
\lambda =\pm \sqrt{2\gamma }.
\label{eig}
\end{eqnarray} 
Hence, the method that we use to derive the eigenvalues $\lambda$
(which fully determine the crucial issue of stability of the solution
for small $C$) consists of a perturbative expansion of the solution
from the AC limit 
\begin{eqnarray}
\phi _{\mathbf{n}}=\phi _{\mathbf{n}}^{(0)}+C \phi
_{\mathbf{n}}^{(1)}+\dots, 
\label{expansion}
\end{eqnarray}
which allows us to derive the principal
bifurcation conditions for a specific configuration and therefore
infer its linear stability properties through the eigenvalues of
$\mathcal{M}$ and their connection to the linearization eigenvalues
$\lambda$. Recall that a nonzero real part of {\it any} 
eigenvalue is a necessary
and sufficient condition for an exponential instability in Hamiltonian
systems, such as the one considered herein.

\section{Comparison of Analytical and Numerical Results}

\subsection{General Terminology} 
\begin{figure}[tbp]
\begin{center}
\hskip-0.15cm
\begin{tabular}{cc}
\includegraphics[height=6cm,width=6cm]{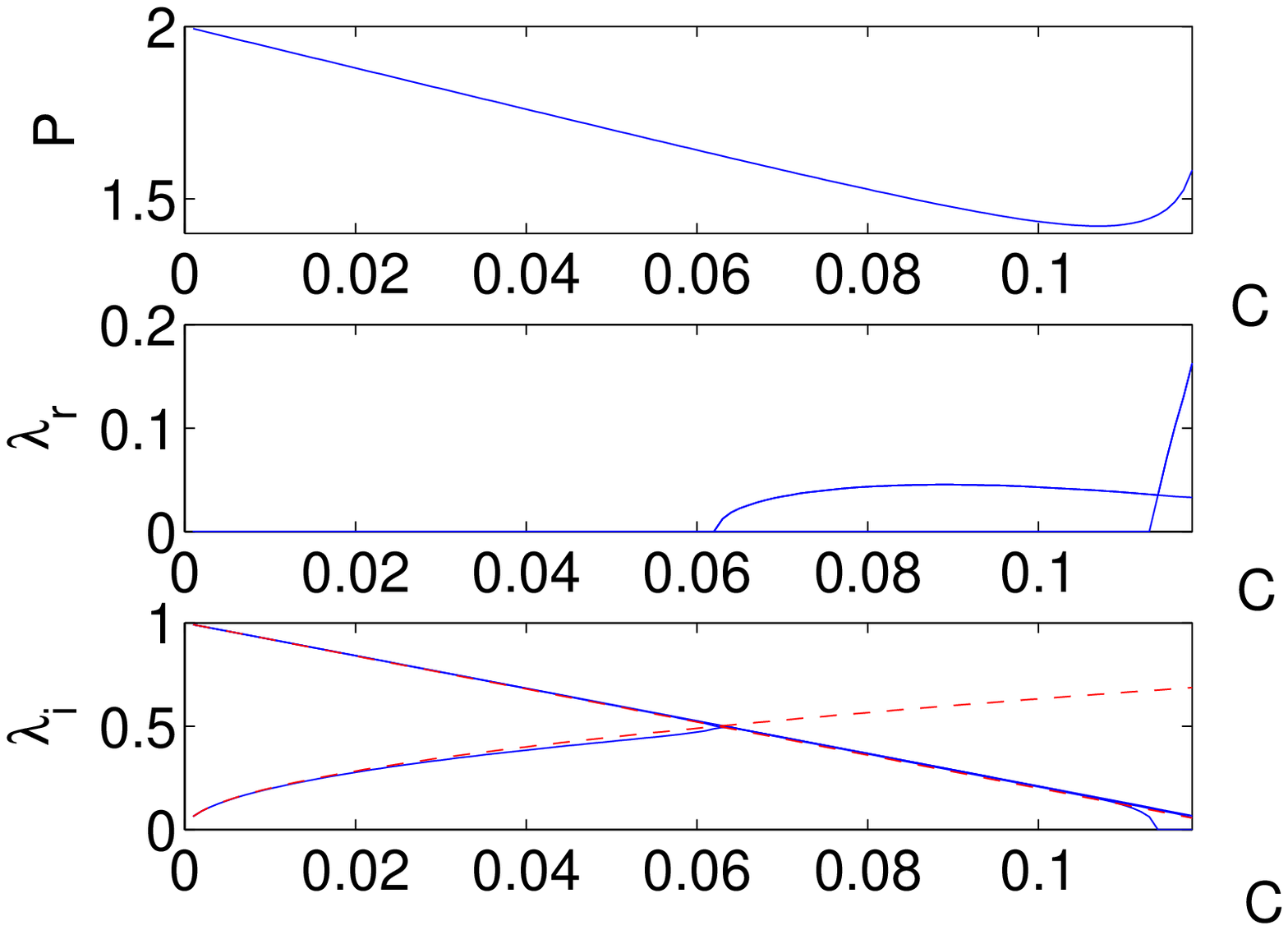}
\includegraphics[height=6cm,width=6cm]{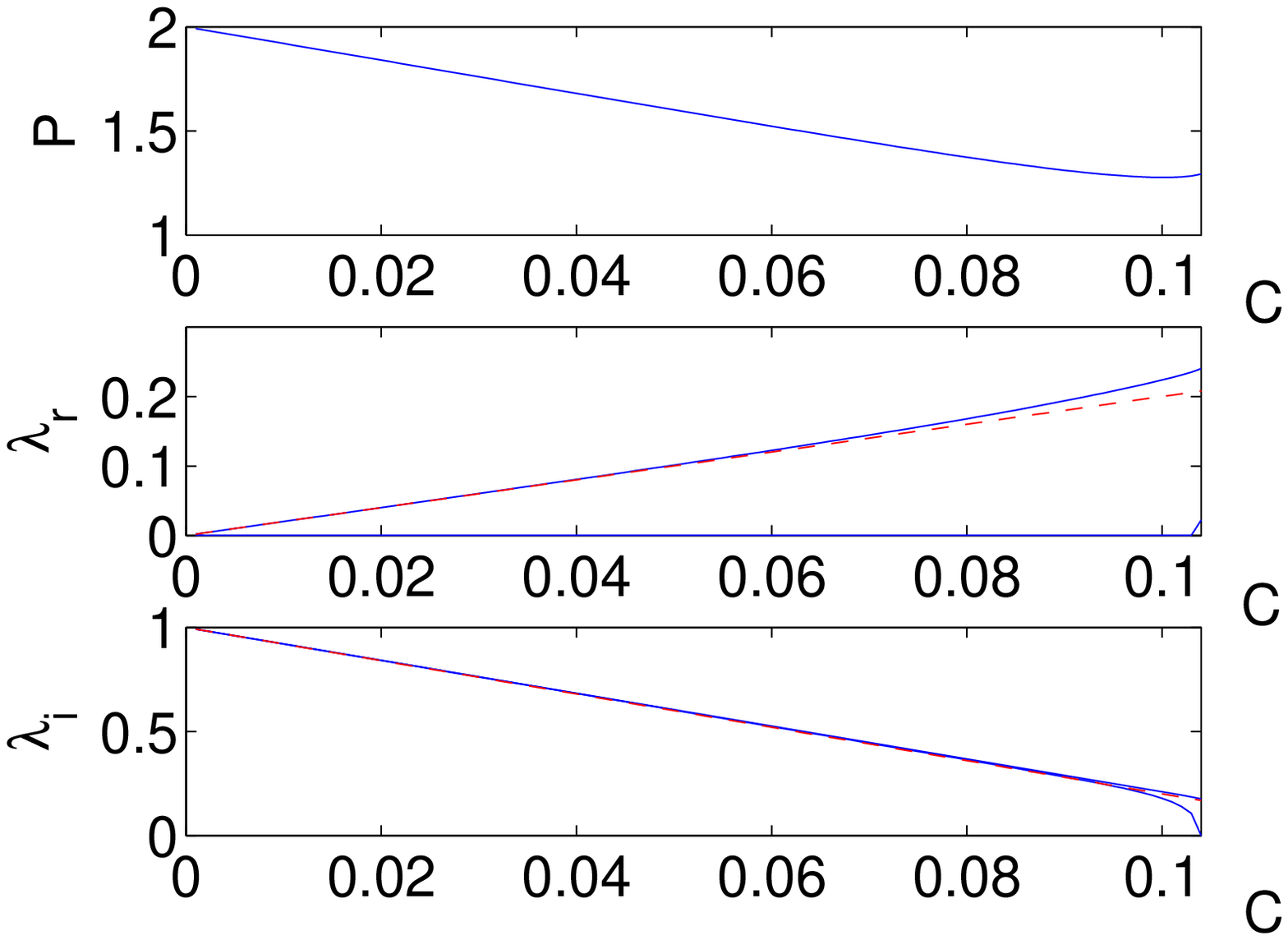} &  \\
\includegraphics[height=6cm,width=6cm]{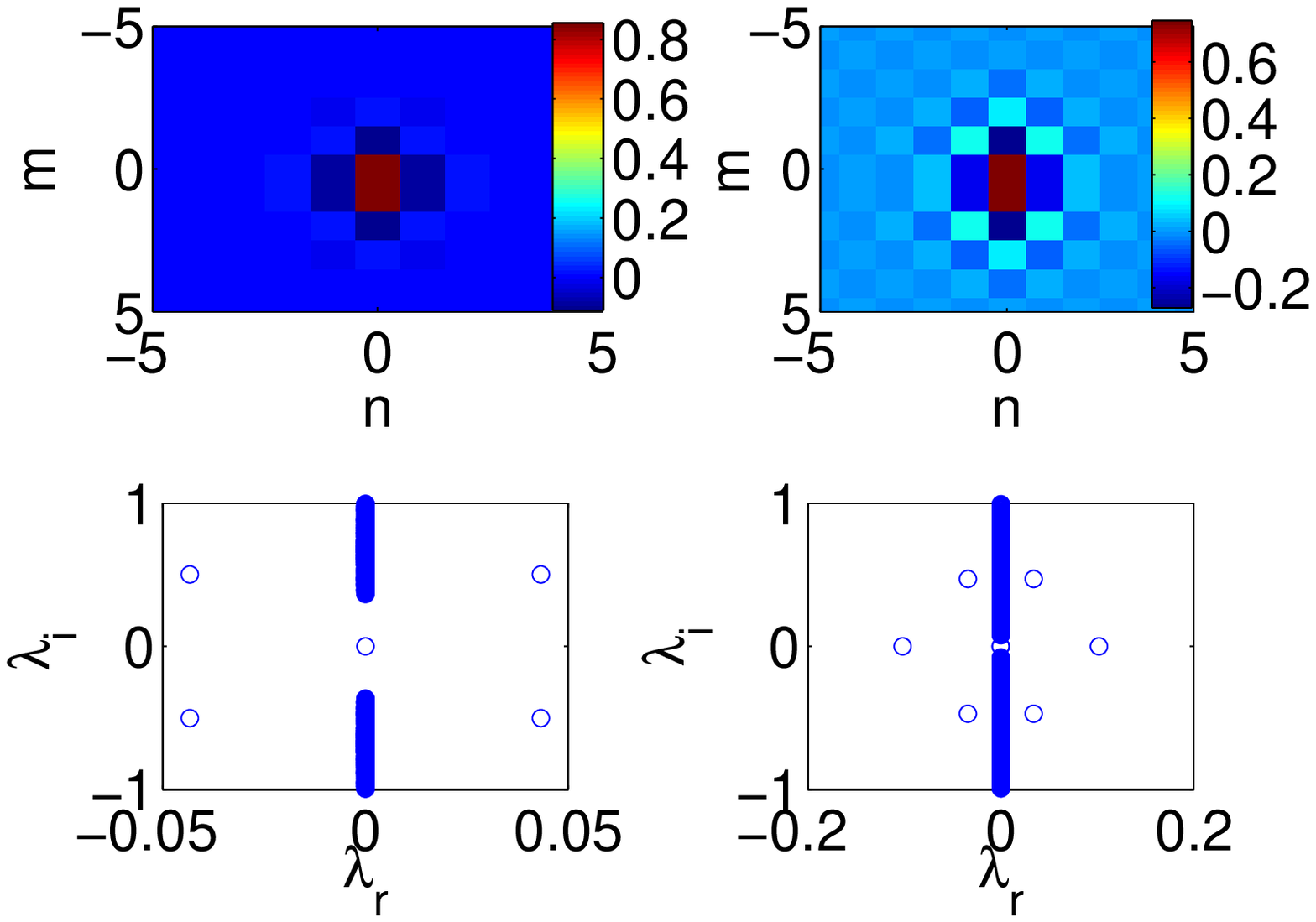}
\includegraphics[height=6cm,width=6cm]{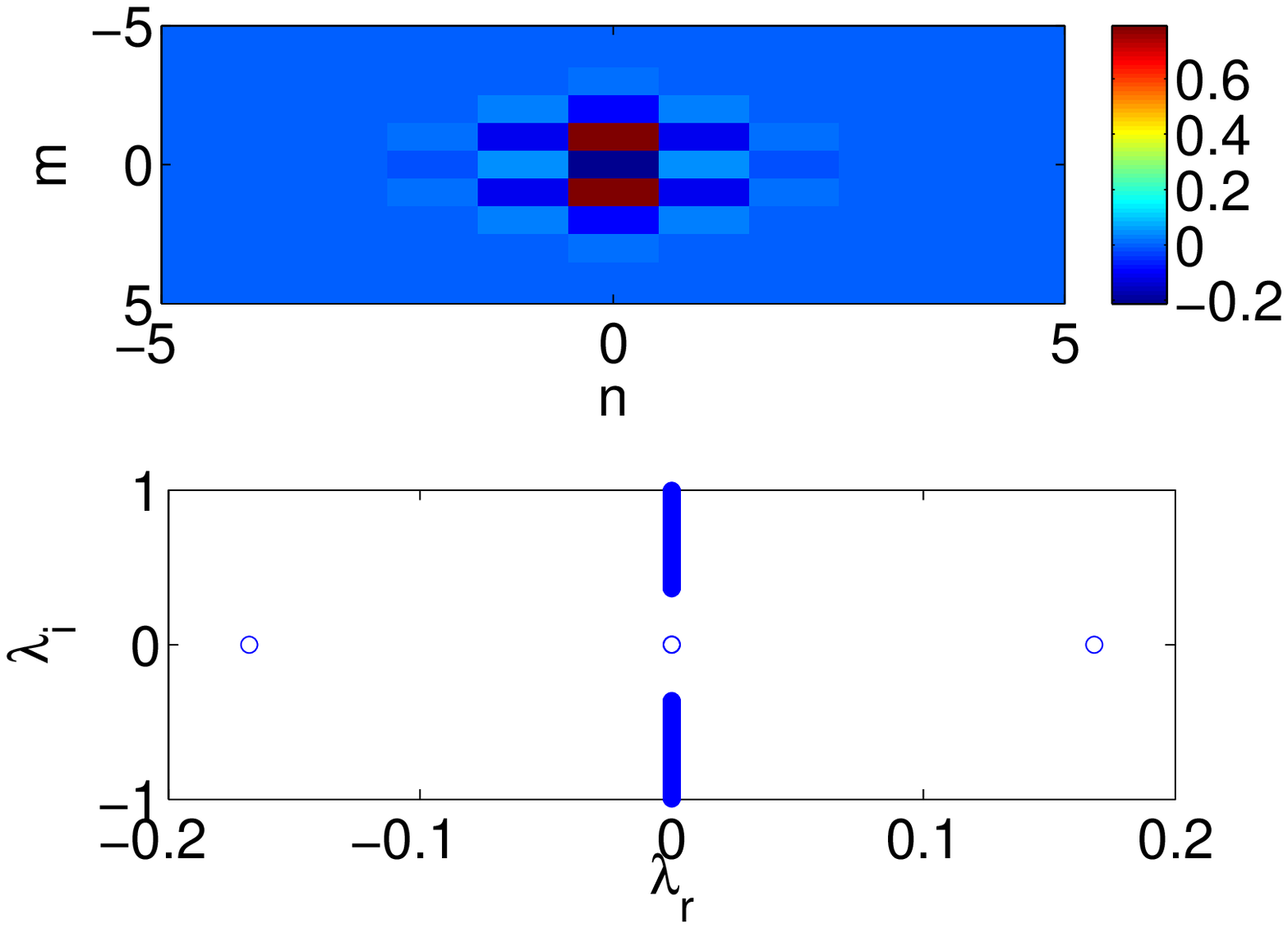} &
\end{tabular}\end{center}
\par
\vskip-0.7cm 
\caption{(Color Online) 
The first line of panels shows the power $P$ vs. coupling $C$
for the inter-site (IS), in-phase (IP) mode (left) and on-site (OS),
IP mode (right). The second lines show their maximal real eigenvalues
and the third their first few imaginary eigenvalues.
The solid (blue) lines illustrate the numerical results, while
the dashed (red) lines the analytical ones. 
The fourth and fifth panels show the contour plot of
the mode profile (fourth panel) and the corresponding spectral plane
of eigenvalues $\lambda=\lambda_r + i \lambda_i$ (fifth panel);
The left two panels are for the IS-IP mode for $C=0.08$ and $C=0.116$
respectively. The right panel shows the OS-IP mode for $C=0.08$.}
\label{fig1}
\end{figure}

\begin{figure}[tbp]
\begin{center}
\hskip-0.15cm
\begin{tabular}{cc}
\includegraphics[height=6cm,width=6cm]{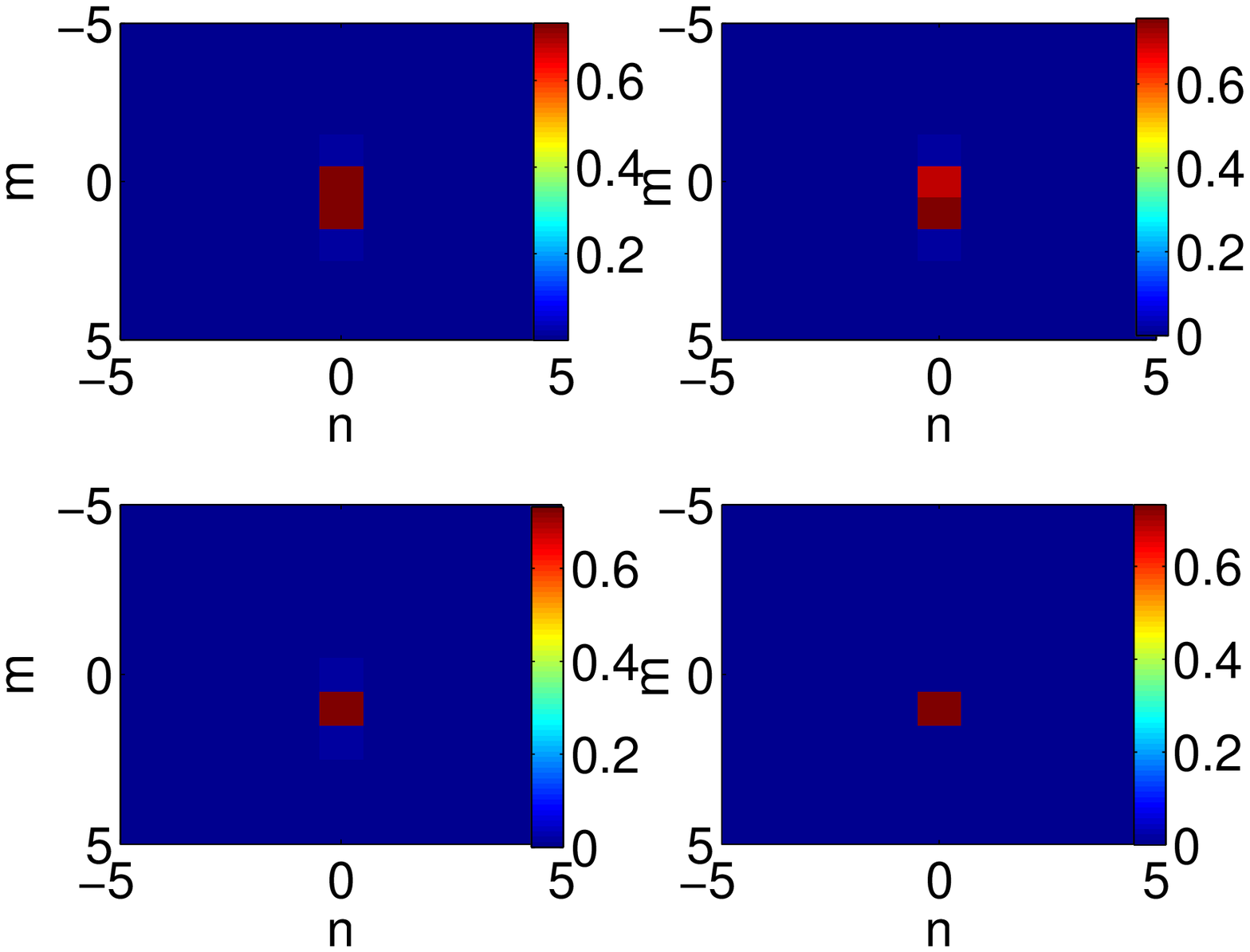}
\includegraphics[height=6cm,width=6cm]{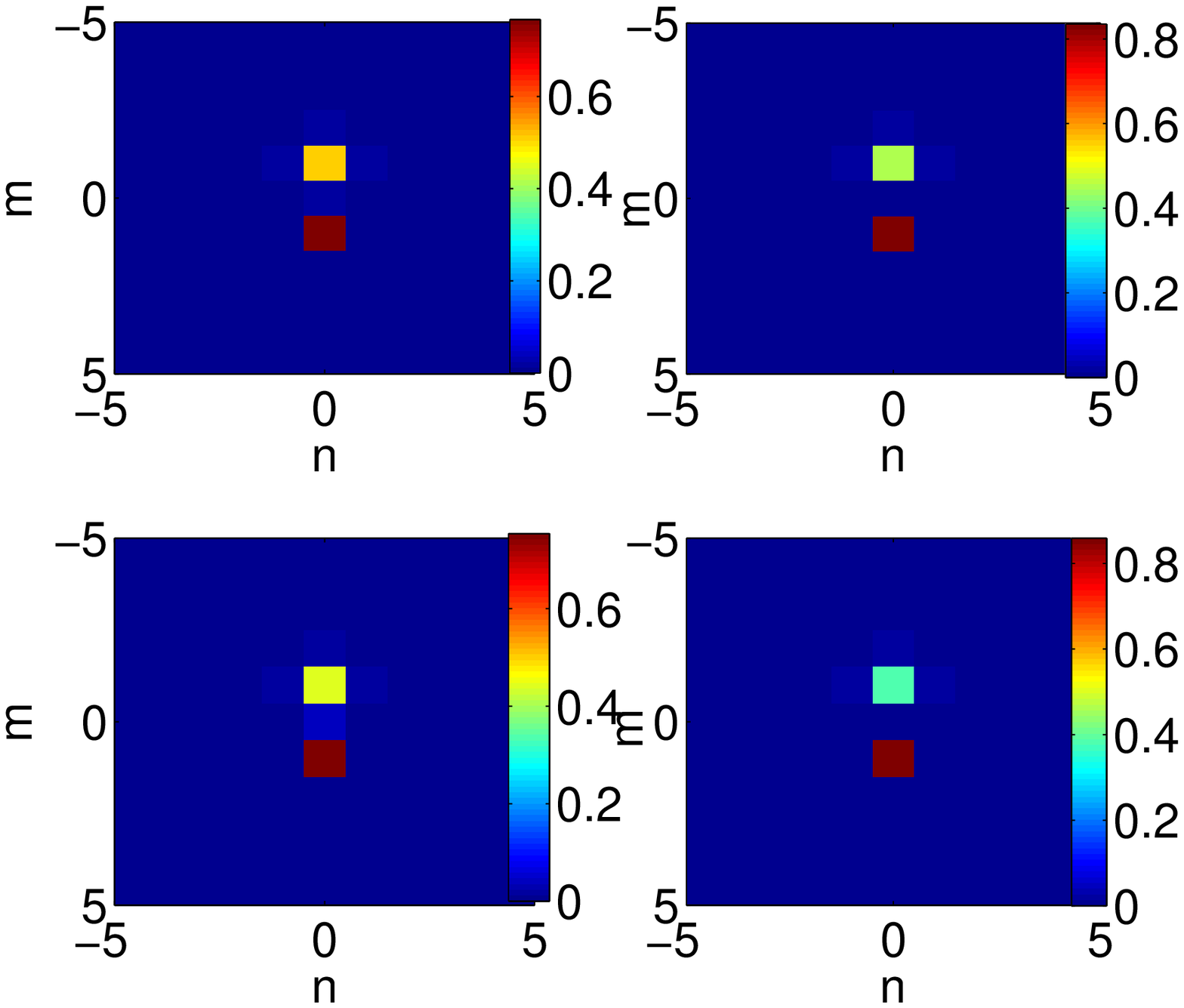} &  \\
\includegraphics[height=6cm,width=6cm]{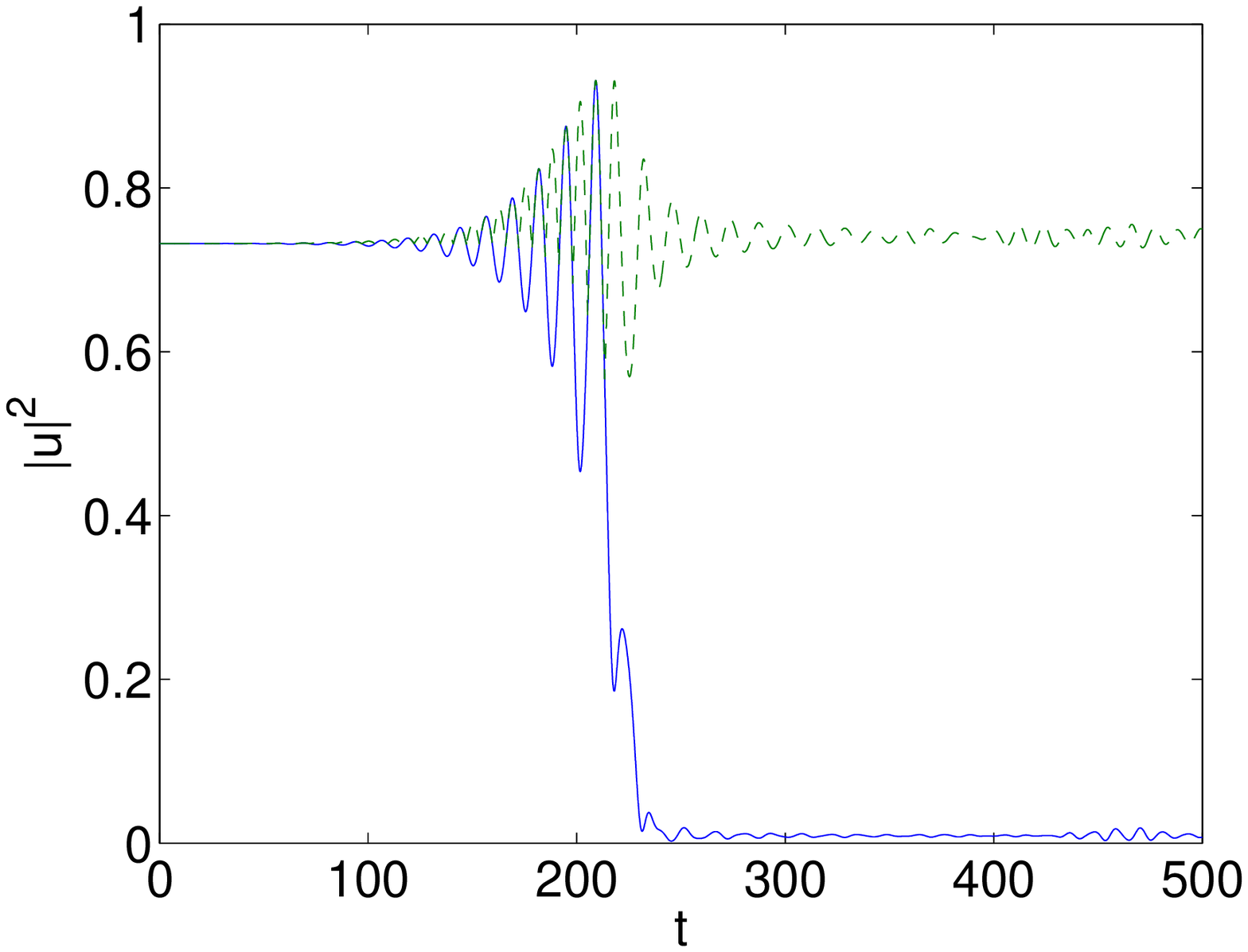}
\includegraphics[height=6cm,width=6cm]{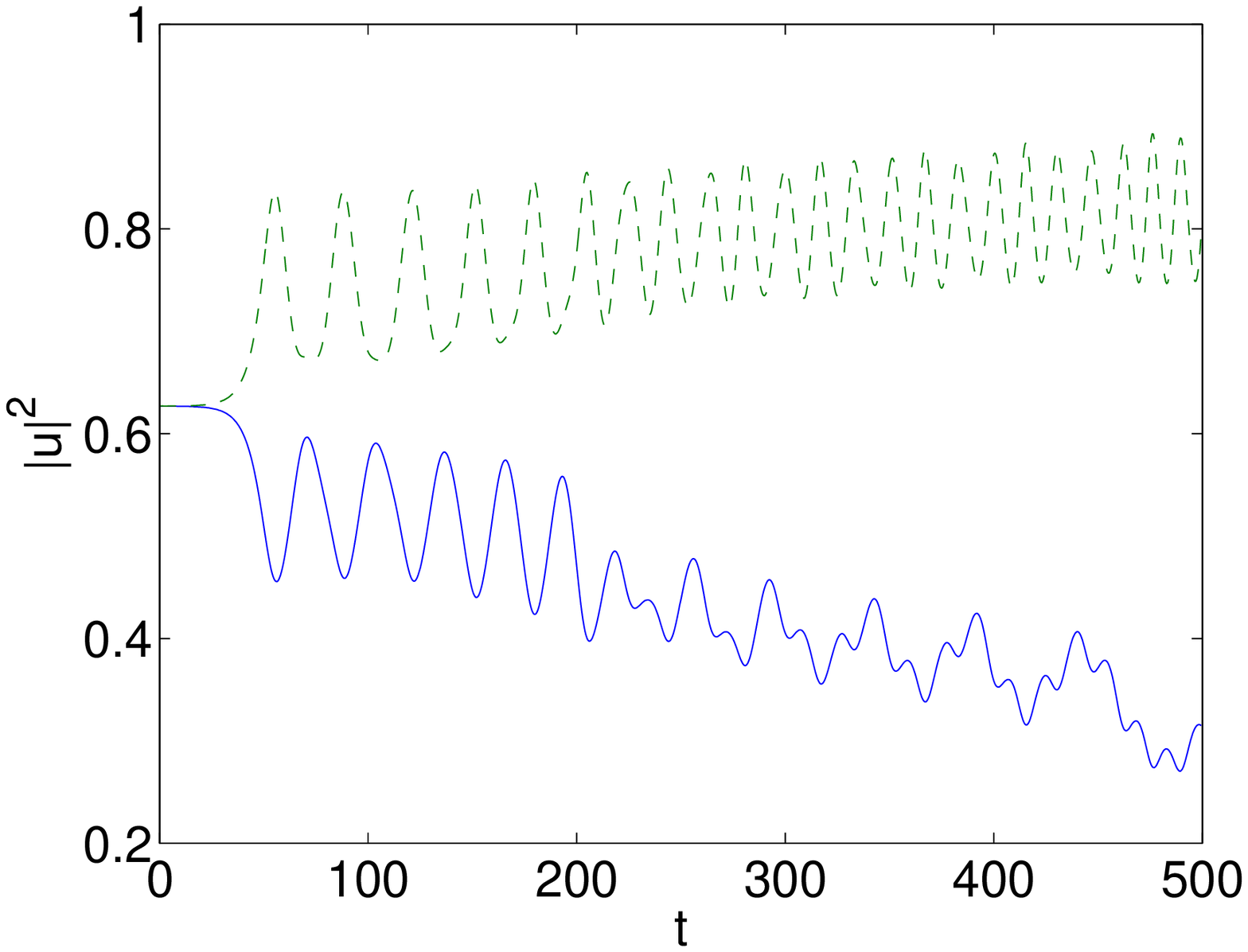} &
\end{tabular}
\end{center}
\par
\vskip-0.7cm 
\caption{(Color Online) The four panels at the top left corner show the contour
plots of the dynamical evolution of the unstable inter-site,
in phase (IS-IP) solution for $C=0.08$. The respective times are 
$t=50$ and $t=150$ in the top and $t=250$ and $t=350$ in
the second row. The panel at the bottom left shows the dynamical
evolution of the square modulus of the 
principal two sites participating in the IS-IP
solution as a function of time. From both of the above, 
it is clear that the configuration relaxes into a single site soliton.
The right panels show the same features for the on-site, in-phase
(OS-IP) solution, which also relaxes (but more slowly) 
into a single-site configuration.}
\label{fig1a}
\end{figure}

We start with 
some general terminology
that we will use in this section. The designation in-phase
(IP) will be used for two sites such that their relative phase difference
is $0$, while out-of-phase (OP) will signify that it is $\pi$. Furthermore,
on-site (OS) will mean that the center of the configuration is on an
empty lattice
site (between the excited ones), while inter-site (IS) will signify that 
the center is located between the excited lattice
sites (and no empty site exists between them). 
For all modes, in the figures below, we show their power $P=\sum
|u_{\mathbf{n}}|^2$ as a function of the coupling strength $C$, as
well as the real and imaginary parts of the key eigenvalues (the
ones determining the stability of the configuration). 
We start with the dipole configuration (consisting primarily of two
lattice sites; see Figs. \ref{fig1}-\ref{fig1c}). We also examine 
the more complex quadrupole (see Figs. \ref{fig2}-\ref{fig2c})
and vortex (see Figs. \ref{fig3}-\ref{fig3a})
configurations. In all the cases, we 
offer typical examples of the mode profiles and
stability for select values of $C$. When the configurations are
found to be unstable, we also give a typical example of the
instability evolution, for a relevant value of the coupling strength.
Another general feature that applies
to all modes is a continuous spectrum band extending for 
$\lambda_i \in [\Lambda-8C, \Lambda]$. This latter trait significantly
affects the stability intervals of the structures in 
comparison with their focusing counterparts as we will see also below
(since configurations may be stable for small $C$, but not for larger
$C$).

The presentation of the figures will be uniform throughout the
manuscript in that in each pair of figures, we examine two
types of configurations (one in the left column and one in the
right column). The first figure of each pair will have five panels showing
$P$ as a function of $C$, the principal real eigenvalues
(second panel) and imaginary eigenvalues (third panel). In
these plots, the numerical results are shown by the solid (blue)
line, while the analytical results by the dashed (red) line.
The fourth and fifth panels show typical examples of the
relevant configuration (obtained through a fixed point iteration
of the Newton type) and its stability eigenvalues 
(shown through the spectral plane $(\lambda_r,\lambda_i)$
for the eigenvalues $\lambda=\lambda_r + i \lambda_i$).
The accompanying second figure will show the result of a typical
evolution of an unstable mode, perturbed by a random perturbation
of amplitude $10^{-4}$, in order to accelerate the instability evolution.
The four contour plot panels (one set on the left and one on the
right) will display the solution's squared absolute value for four
different values of the evolution variable; the bottom panel will
show the dynamical evolution of the sites chiefly ``participating''
in the solution. A fourth-order Runge-Kutta scheme has been used for
the numerical integration results presented herein.

To facilitate the reader, a summary of the results, encompassing
our main findings reported below is offered in Table 1. The
table summarizes the configurations considered, their
linear stability and the outcome of their dynamical evolution
for appropriate initial conditions in the instability regime. 
Note that if the solutions are unstable for all $C$, they
are denoted as such, while if they are partially stable for
a range of coupling strengths, their interval of stability is
explicitly mentioned. Details of our analytical results and
their connection/comparison with the numerical findings
are offered in the rest of this section.

\begin{table*}[t]
\begin{center}
\begin{tabular}{|l|c|c|c|c|}
\hline
& \multicolumn{2}{|c|}{On-site} & \multicolumn{2}{|c|}{Inter-site} \\
\cline{2-5}
Type & Stability & Instability Outcome & Stability & Instability Outcome \\ \hline
In-phase Dipole & Unstable & 1-Site Pulse & $C<0.064$ & 1-Site Pulse \\ \hline
Out-of-phase Dipole & $C<0.092$ & Decay & Unstable & 1-Site Pulse \\ \hline
In-phase Quadrupole & Unstable & Breathing Behavior & $C<0.047$ & 1-Site Pulse 
\\ \hline
Out-of-phase Quadrupole & $C<0.08$ & 1-Site Pulse & Unstable & 2-Site Mode \\ \hline
Vortex & $C<0.095$ & 1-Site Pulse & $C<0.095$ & 1-Site Pulse \\ \hline
\end{tabular}  \end{center}
\caption{Summary of the stability results for all the configurations
presented below. For partially stable (near the anti-continuum limit)
solutions their interval of stability (for $\Lambda=1$) is given.
In each case, the outcome of the instability evolution for the parameters 
and initial conditions considered below is also mentioned.} \label{table1}
\end{table*}

\subsection{Dipole Configurations}

\subsubsection{Inter-site, In-Phase Mode}
Figures \ref{fig1}-\ref{fig1a} encompass our results for the
two types of IP dipole solutions
(i.e., initialized at the AC limit with two in-phase excited sites). 
The 
IS-IP mode of the left panels 
is theoretically found to possess 1 imaginary eigenvalue
pair (and, hence, is stable for small $C$)
\begin{eqnarray}
\lambda \approx \pm 2 \sqrt{C} i.
\label{ISIP}
\end{eqnarray}
The collision with the continuous
spectrum described above 
causes the mode to become unstable for sufficiently large
$C$; the theoretically predicted instability
threshold (obtained by equating the eigenvalue of 
Eq. (\ref{ISIP}) with the lower edge of the phonon 
band located at $\Lambda-8 C$) is $C=0.0625$, the 
numerically found one is $C \approx 0.064$. Additional
instability may ensue when the monotonicity of the $P$ vs. $C$ curve
changes (we have found this to be a general feature of the defocusing 
branches). The fourth and fifth panels show the mode and its
linearization eigenvalues for $C=0.08$ and $C=0.116$. 
In fact, the dynamical evolution of the mode is demonstrated
for the case of $C=0.08$, illustrating that only one of the two
sites eventually persists, after the demonstrably oscillatory
instability destroys the configuration for $t>100$.

\subsubsection{On-site, In-Phase Mode}
The OS-IP mode of the right panels of Figs. \ref{fig1}-\ref{fig1a}
is always unstable due to a real pair, theoretically found to be 
\begin{eqnarray}
\lambda \approx \pm 2 C,
\label{OSIP}
\end{eqnarray}
for small $C$. Notice once again the remarkable accuracy
of this theoretical prediction, in comparison with the numerically
obtained eigenvalue.
The fourth and fifth right panels of Fig. \ref{fig1} show the 
mode and its stability for $C=0.08$. Its dynamical evolution in the
right column of Fig. \ref{fig1a} shows its slow disintegration
into a single-site solitary wave.

\subsubsection{Inter-site, Out-of-phase Mode}

\begin{figure}[tbp]
\begin{center}
\hskip-0.15cm
\begin{tabular}{cc}
\includegraphics[height=6cm,width=6cm]{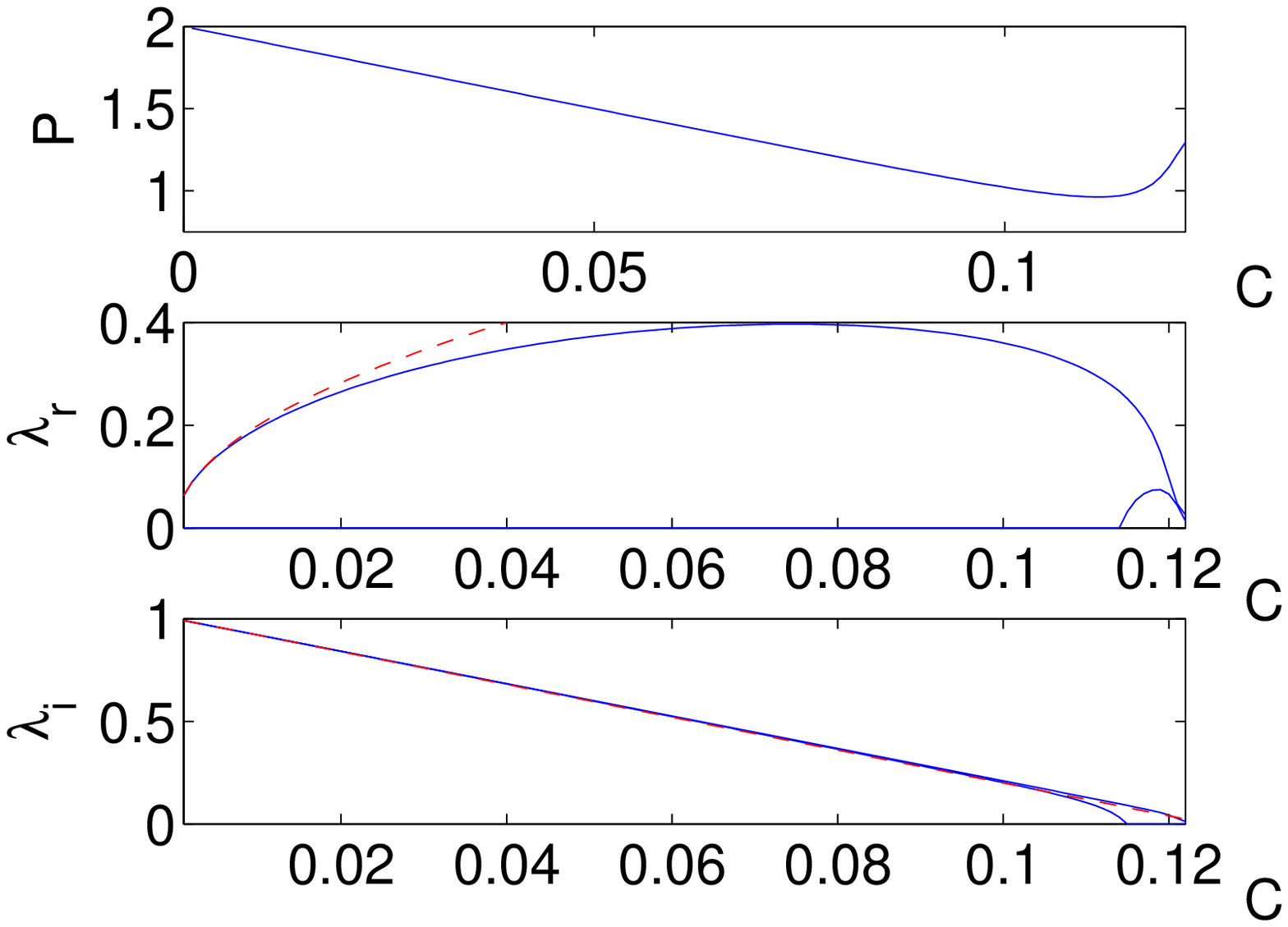}
\includegraphics[height=6cm,width=6cm]{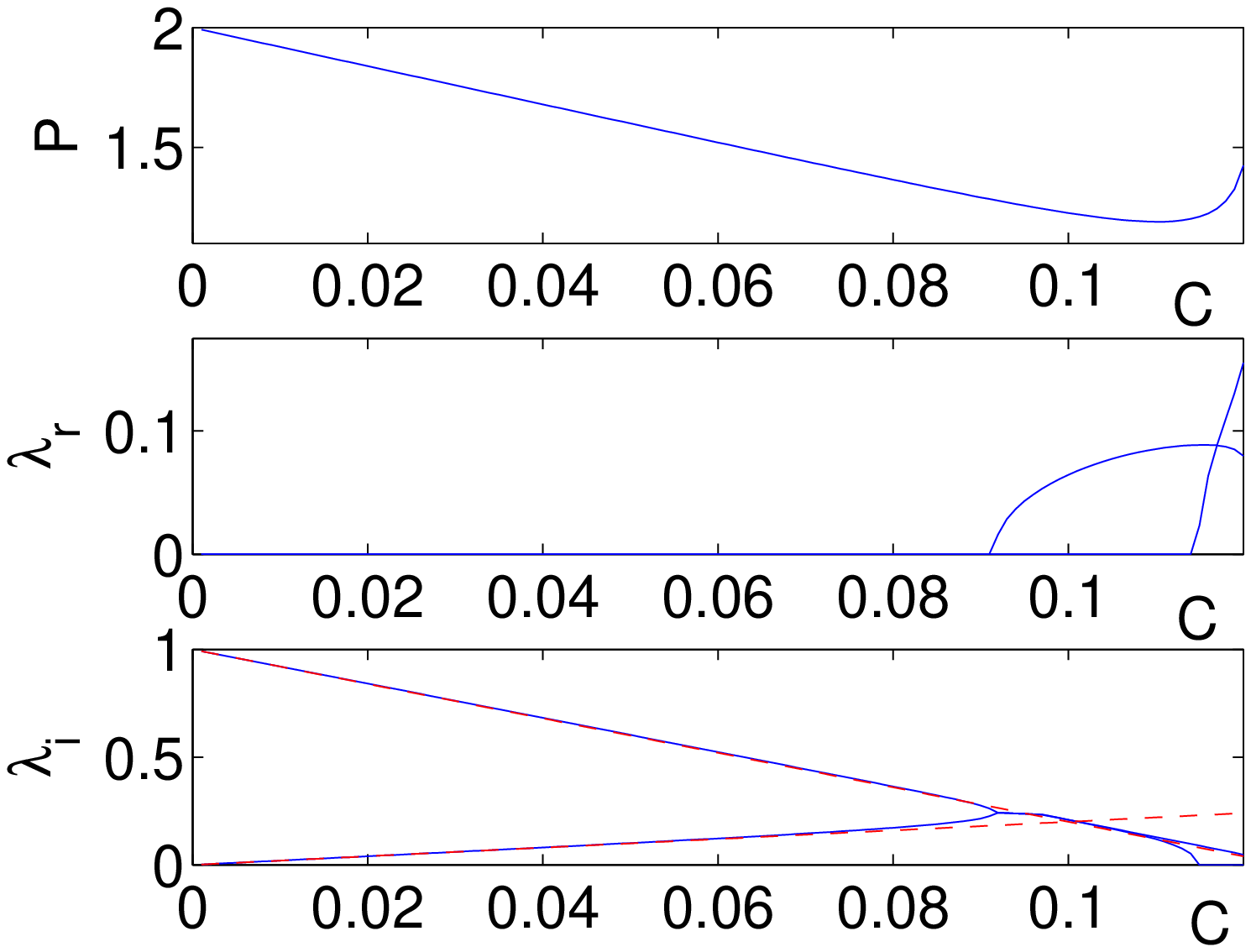} &  \\
\includegraphics[height=6cm,width=6cm]{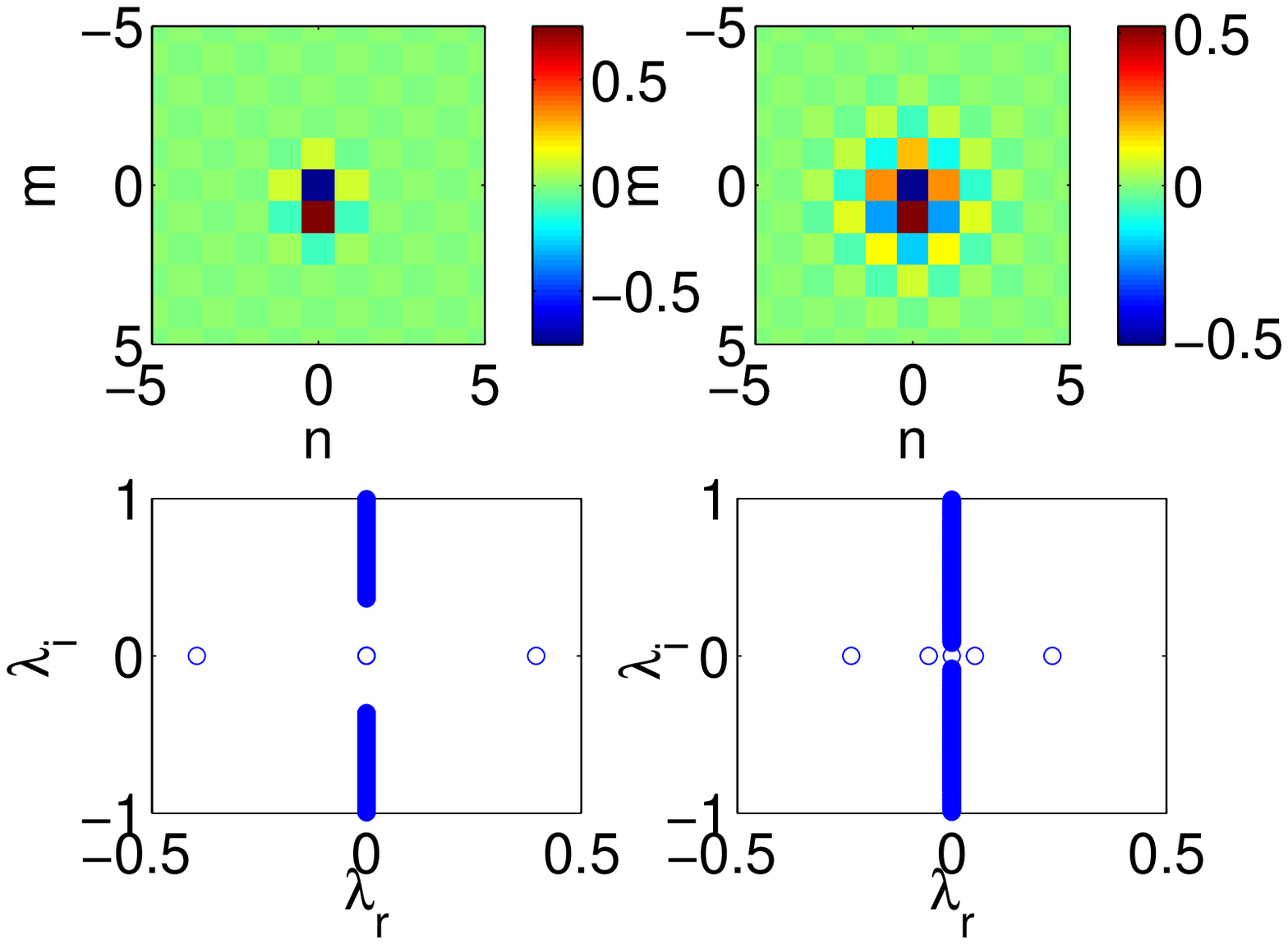}
\includegraphics[height=6cm,width=6cm]{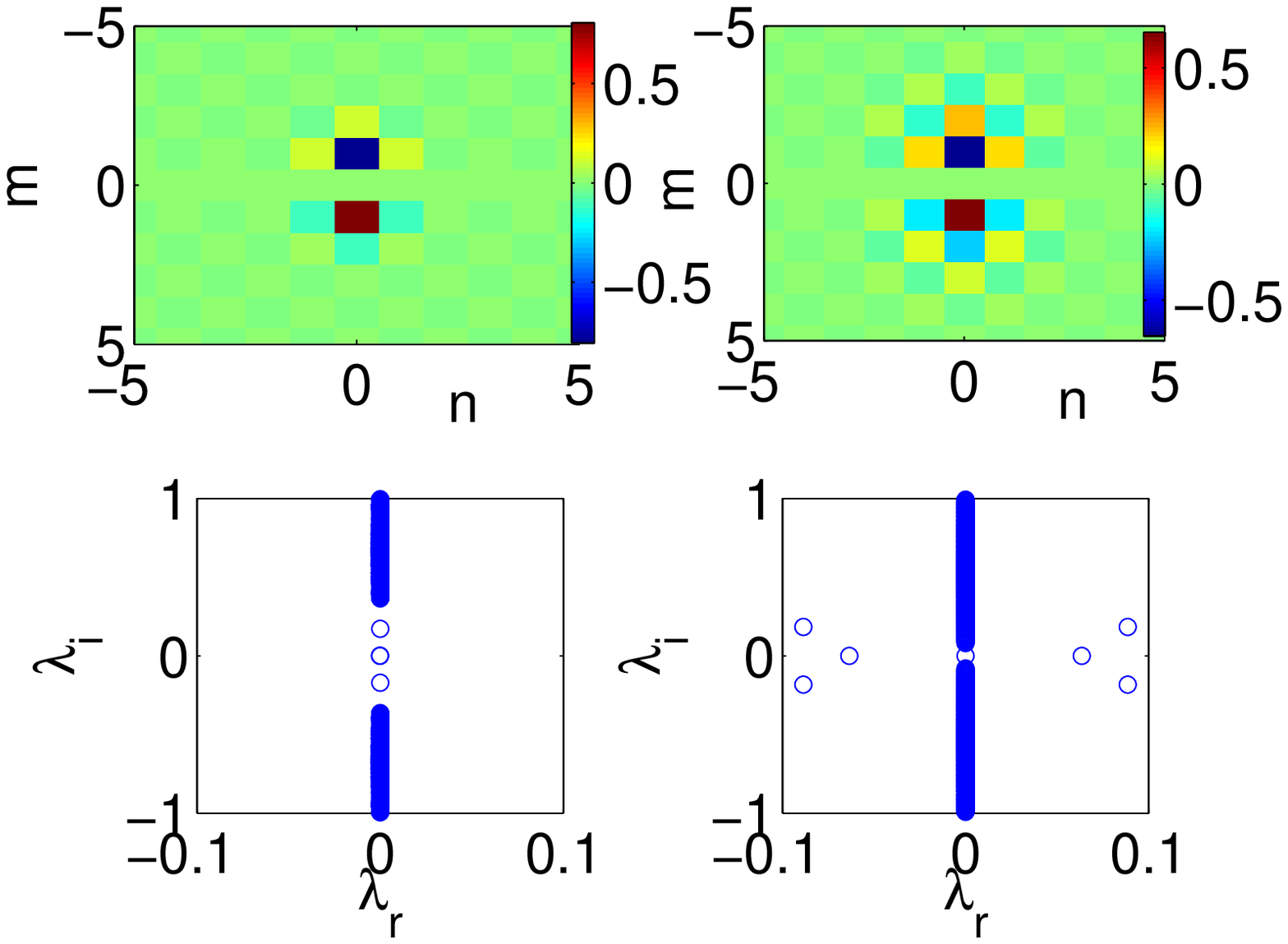} &
\end{tabular}
\end{center}
\par
\vskip-0.7cm 
\caption{(Color Online) Similar to Fig. \ref{fig1}, but now
for the inter-site, out-of-phase (IS-OP) mode (left panels)
and for the on-site, out-of-phase mode (OS-OP). The fourth
and fifth rows of panels are for $C=0.08$ and for
$C=0.116$ in both cases.}
\label{fig1b}
\end{figure}

\begin{figure}[tbp]
\begin{center}
\hskip-0.15cm
\begin{tabular}{cc}
\includegraphics[height=6cm,width=6cm]{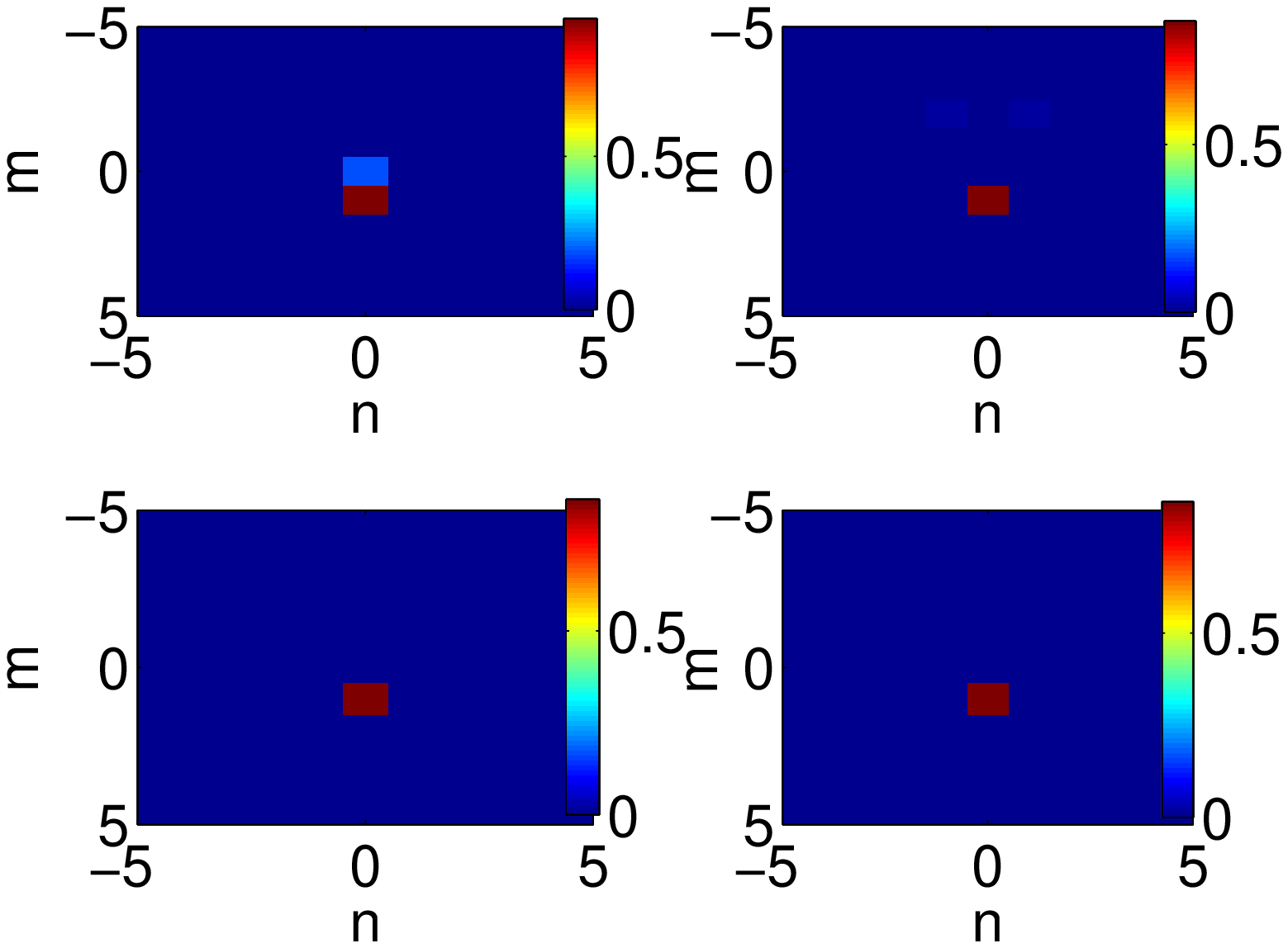}
\includegraphics[height=6cm,width=6cm]{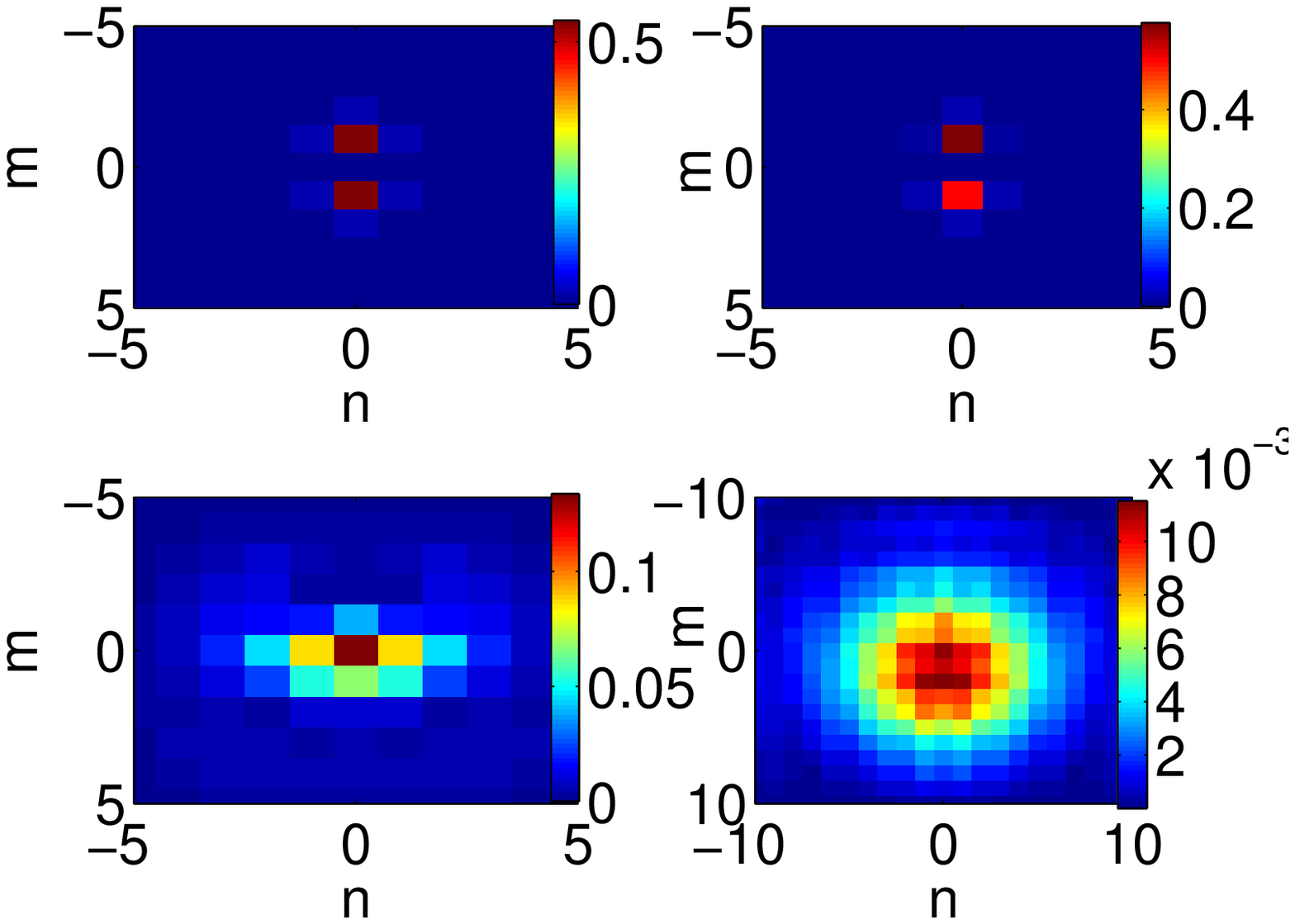} &  \\
\includegraphics[height=6cm,width=6cm]{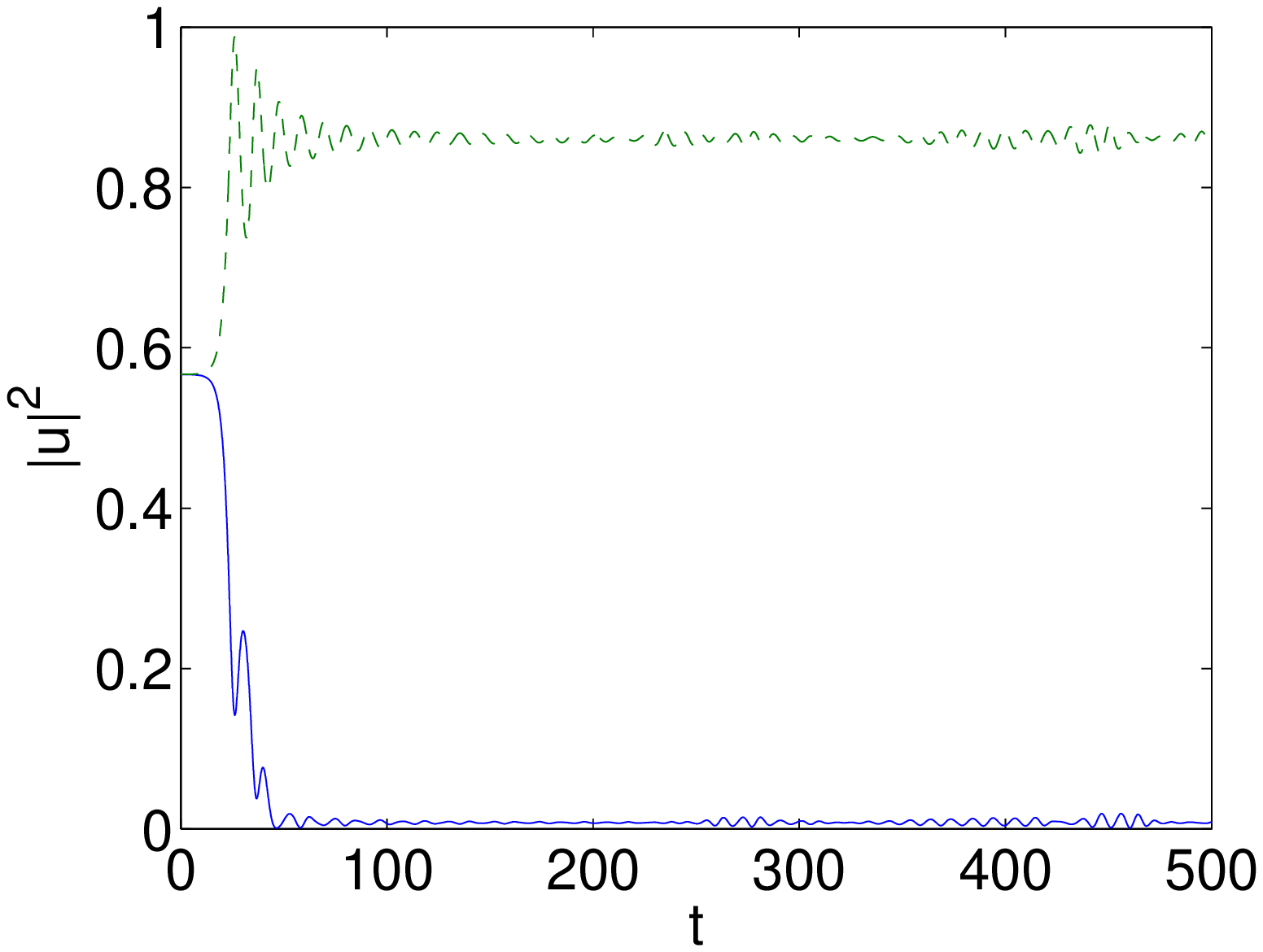}
\includegraphics[height=6cm,width=6cm]{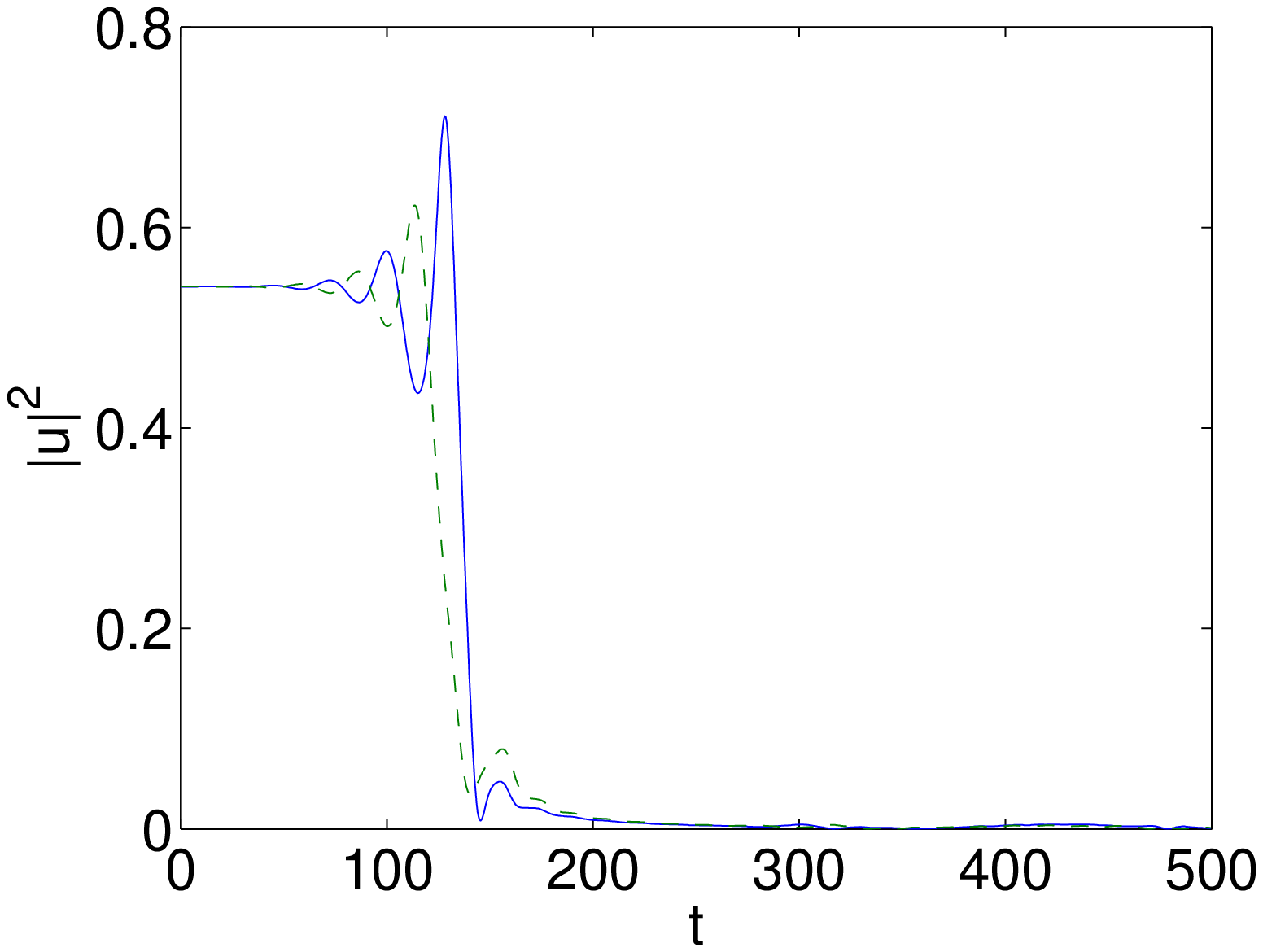} &
\end{tabular}
\end{center}
\par
\vskip-0.7cm 
\caption{(Color Online) Same as in Fig. \ref{fig1a}, but now for the IS-OP mode
(left panels) and the OS-OP mode (right panels). The former shown
at times $t=25$, $t=50$ (top row), $t=150$ and $t=250$ (second row), 
in the case of
$C=0.08$ finally results into a single-site configuration (as is
also illustrated by the bottom panel showing the two principal
sites participating in the mode). The latter, shown for $C=0.1$, 
at times $t=50$, $t=100$ (top row), $t=150$ and $t=200$ (second row)
is entirely destroyed by
the instability resulting into small amplitude excitations.}
\label{fig1c}
\end{figure}

Figures \ref{fig1b}-\ref{fig1c} illustrate the two dipole, out-of-phase
modes.
The left panels of the figures correspond to the IS-OP mode;
this one is also immediately unstable (as one departs from the 
anti-continuum limit),
due to a real pair which is 
\begin{eqnarray}
\lambda \approx 2 \sqrt{C},
\label{ISOP}
\end{eqnarray}
 for small $C$. The fourth and fifth panels of Fig. \ref{fig1b}
show the relevant mode for $C=0.08$ and $C=0.116$, showing its
1 and 2 unstable real eigenvalue pairs respectively.
The numerical experiment highlighting the evolution of the
mode for the case of $C=0.08$ is shown in the left panel
of Fig. \ref{fig1c}. Clearly, in this case as well, the 
positive real eigenvalue leads to the growth of one of the
two sites constituting the dipole, and the eventual formation
of a single-site solitary pulse.

\subsubsection{On-site, Out-of-phase Mode}
The right panels of Fig. \ref{fig1b}-\ref{fig1c} show the
OS-OP mode. The stability analysis of this waveform shows
that it
possesses an imaginary
eigenvalue 
\begin{eqnarray}
\lambda \approx 2 C i.
\label{OSOP}
\end{eqnarray}
This leads to an instability upon
collision (occurring 
theoretically for $C= 0.1$, numerically for $C\approx 0.092$)
with the lower edge (located at $\Lambda-8 C$) of the
continuous band of phonon modes. 
The mode is shown for $C=0.08$ and $C=0.116$ in the right panels
of Fig. \ref{fig1b}.
The direct integration
of the unstable solution with $C=0.1$ is shown in the right panels
of Fig. \ref{fig1c}, indicating that in this case 
the mode completely
disappears (because of the oscillatory instability) into 
extended wave, small amplitude radiation.

\subsection{Quadrupole Confirgurations}

\subsubsection{Inter-site, In-phase Mode}

\begin{figure}[tbp]
\begin{center}
\hskip-0.15cm
\begin{tabular}{cc}
\includegraphics[height=6cm,width=6cm]{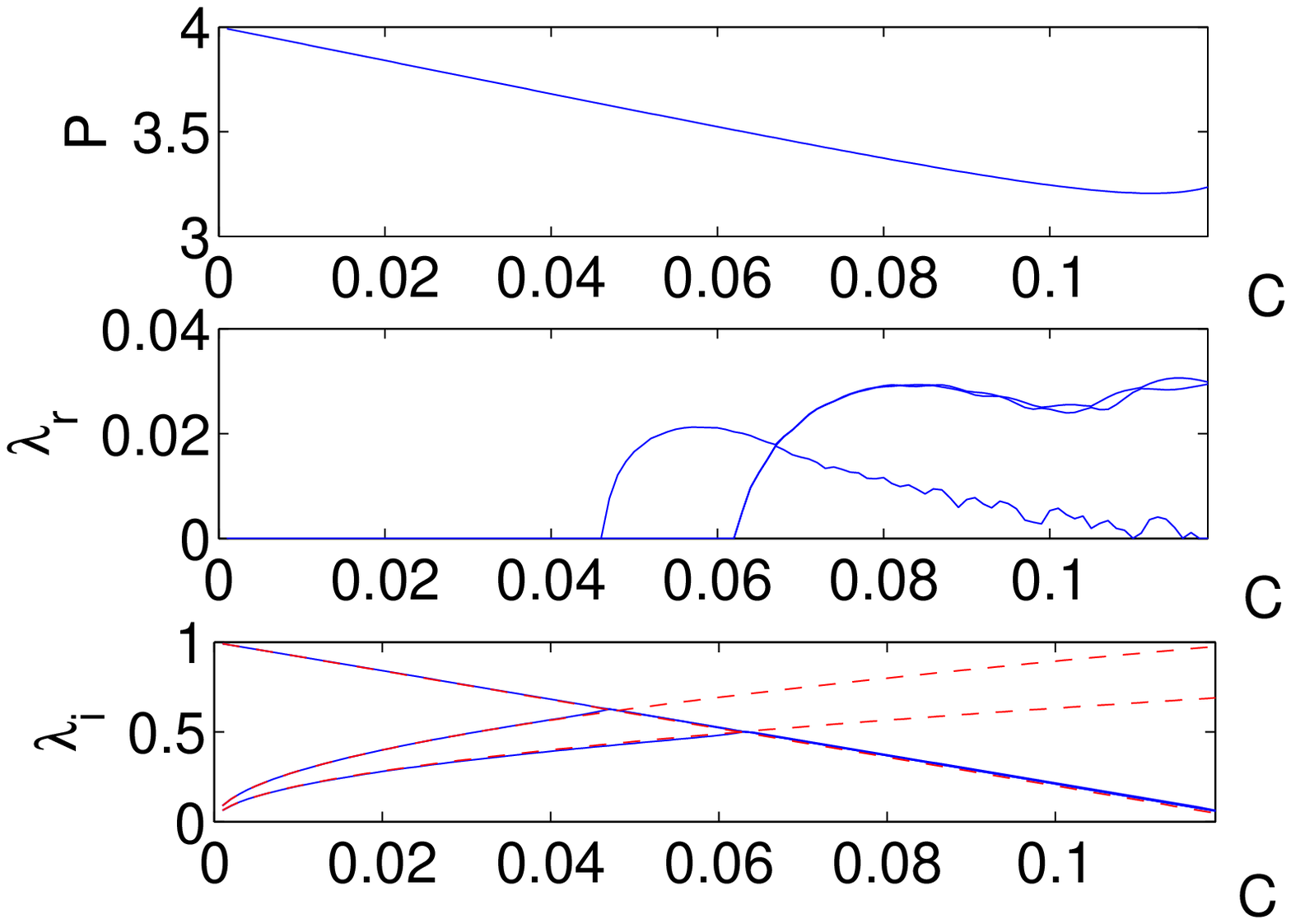}
\includegraphics[height=6cm,width=6cm]{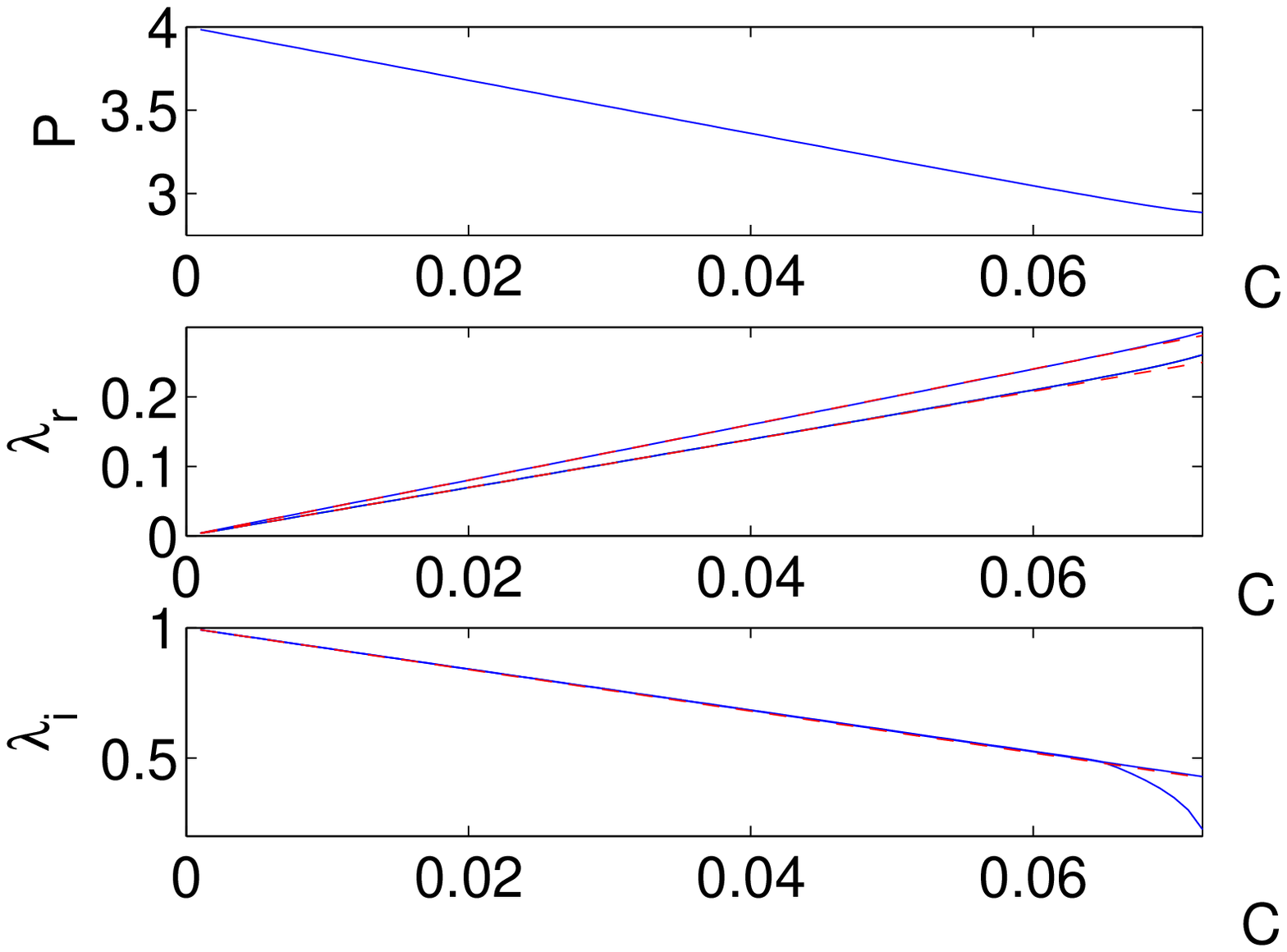} &  \\
\includegraphics[height=6cm,width=6cm]{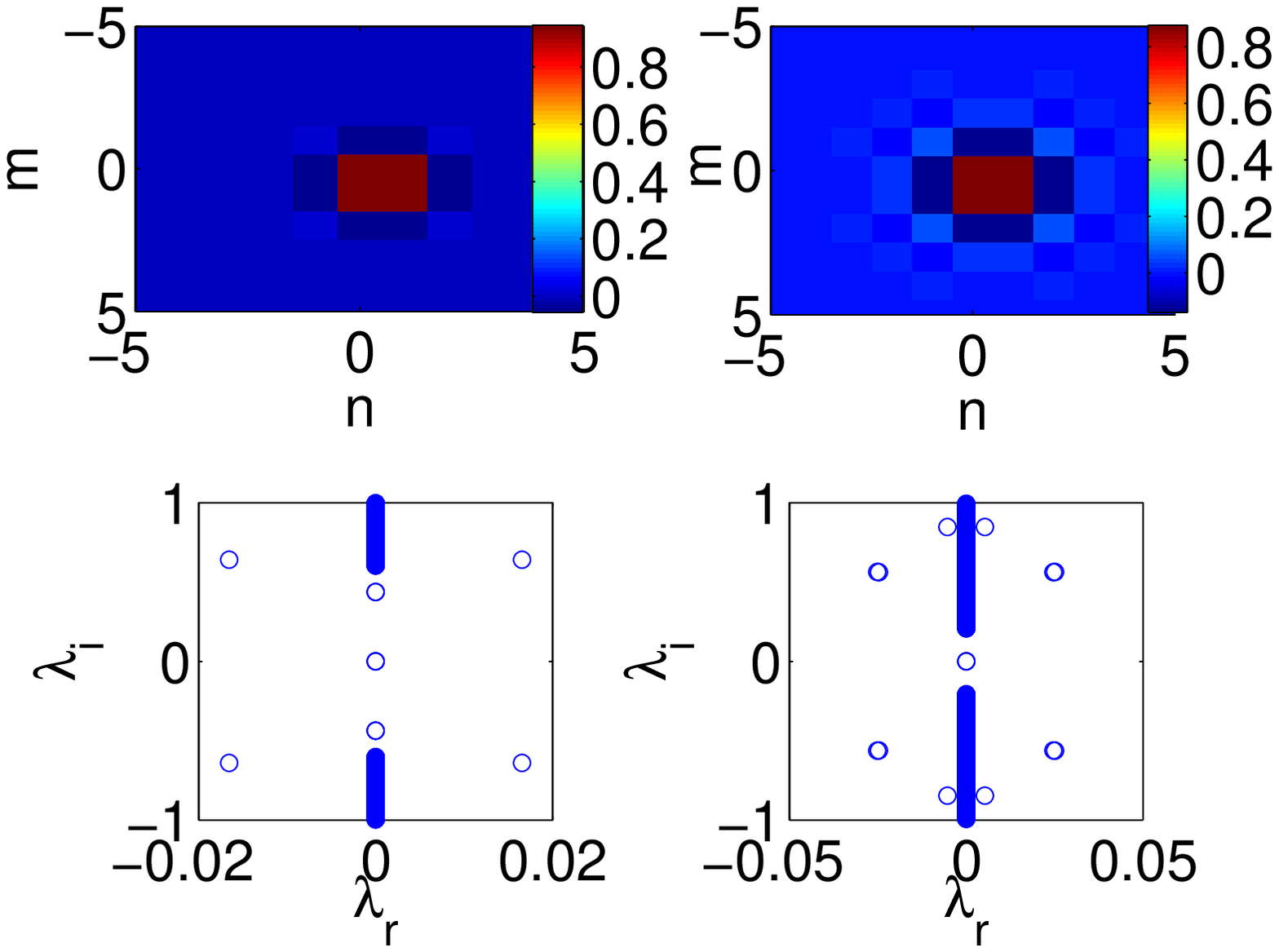}
\includegraphics[height=6cm,width=6cm]{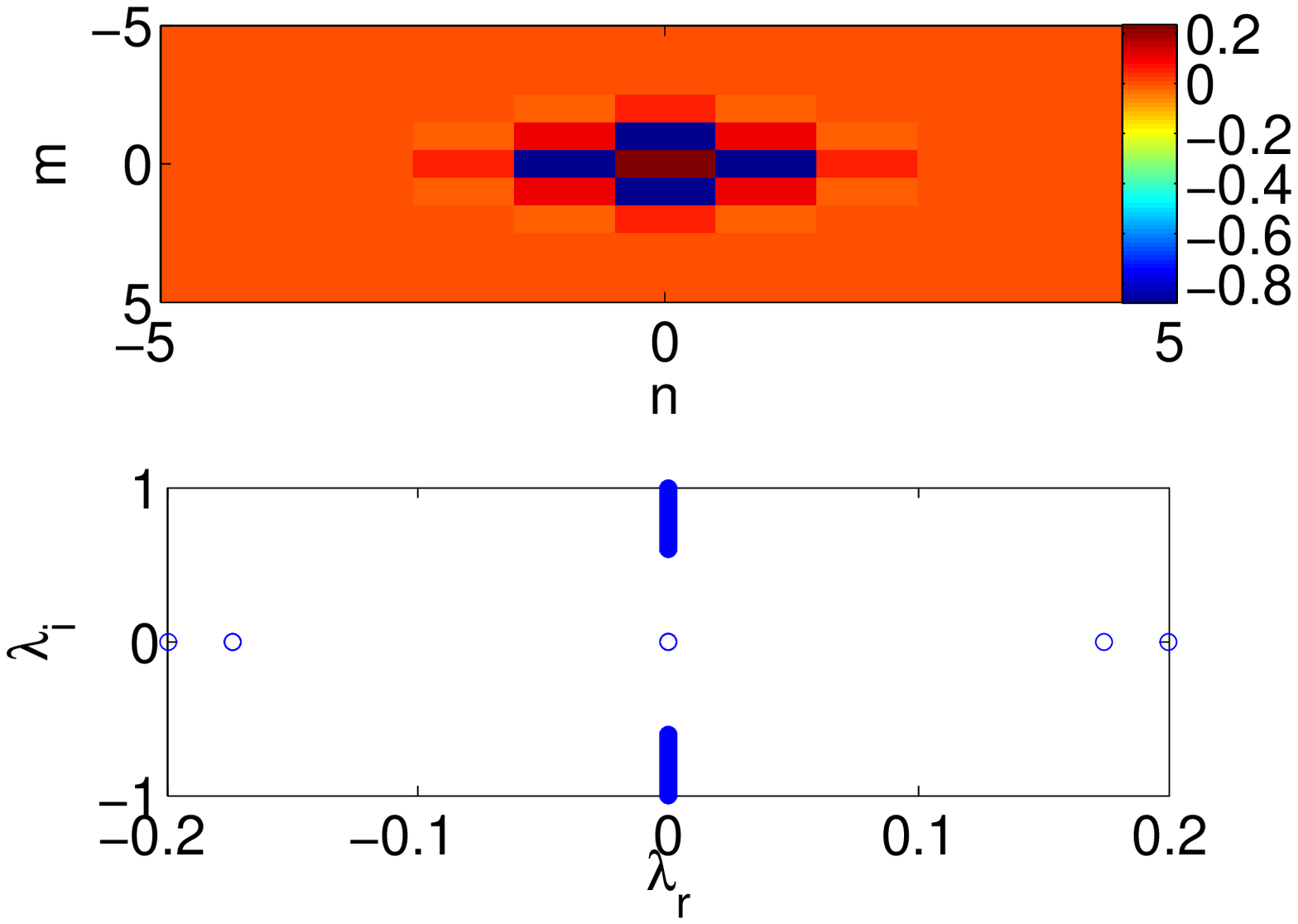} &
\end{tabular}\end{center}
\par
\vskip-0.7cm 
\caption{(Color Online) The top three lines of panels show the same features 
as the corresponding ones of figure \ref{fig1} but now for the quadrupole
IS-IP mode (left) and the quadrupole OS-IP mode (right). 
The contour
plot of the real part of the modes and the spectral plane 
of their linearization eigenvalues are shown in 
the fourth and fifth rows for $C=0.05$ and $C=0.1$ in the case of
the former mode, while  the latter is only shown for $C=0.05$.}
\label{fig2}
\end{figure}

\begin{figure}[tbp]
\begin{center}
\hskip-0.15cm
\begin{tabular}{cc}
\includegraphics[height=6cm,width=6cm]{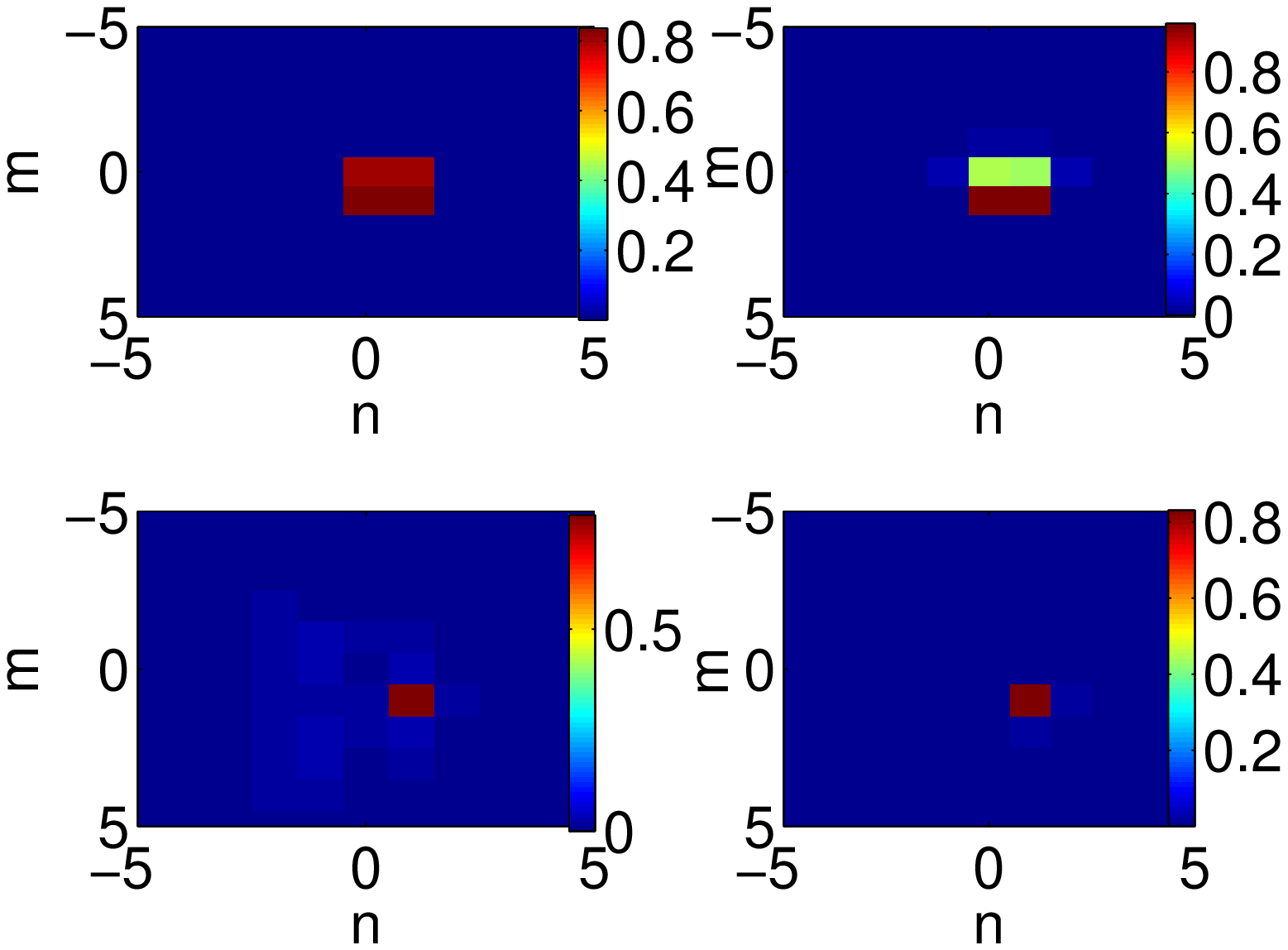}
\includegraphics[height=6cm,width=6cm]{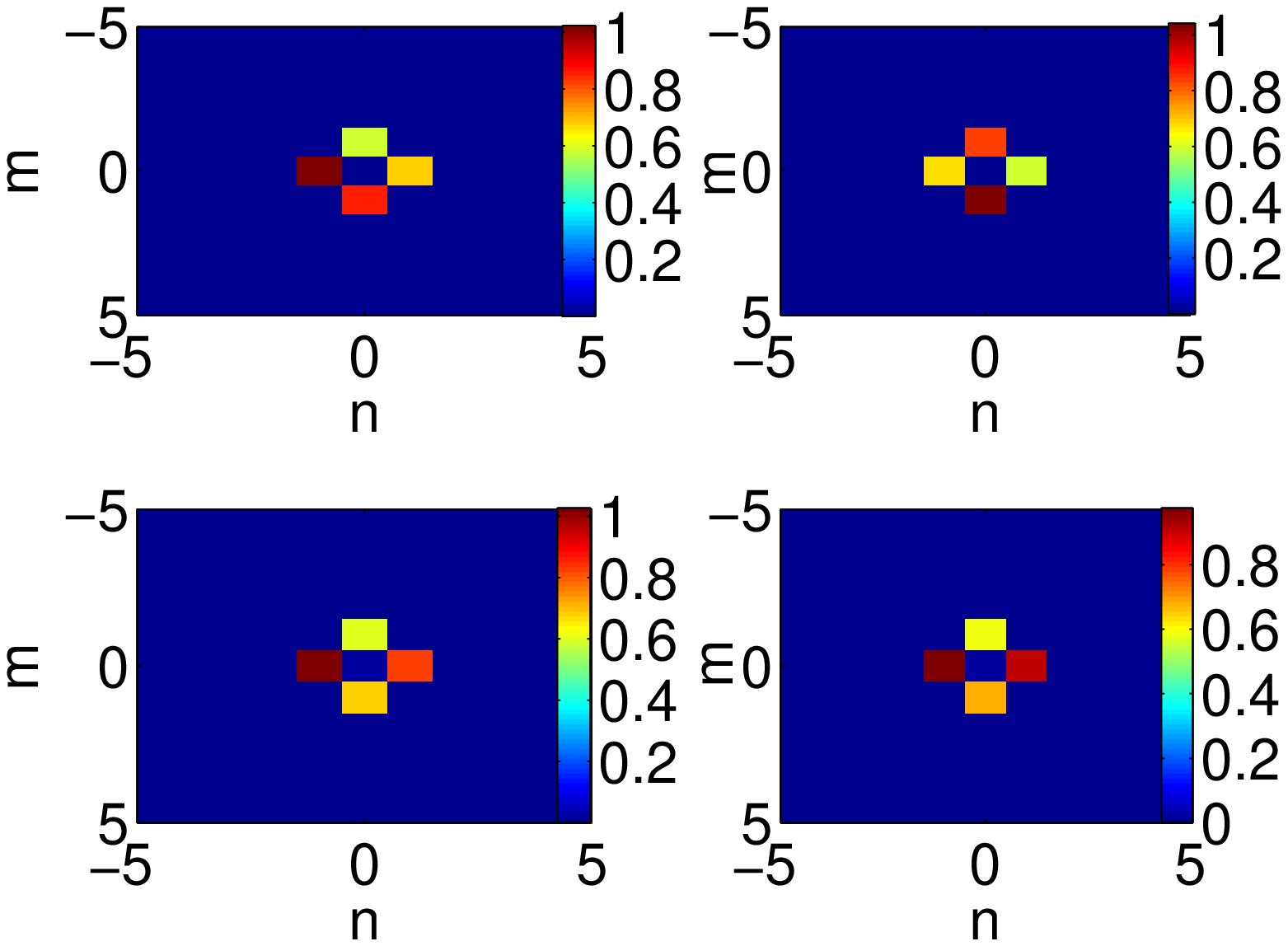} &  \\
\includegraphics[height=6cm,width=6cm]{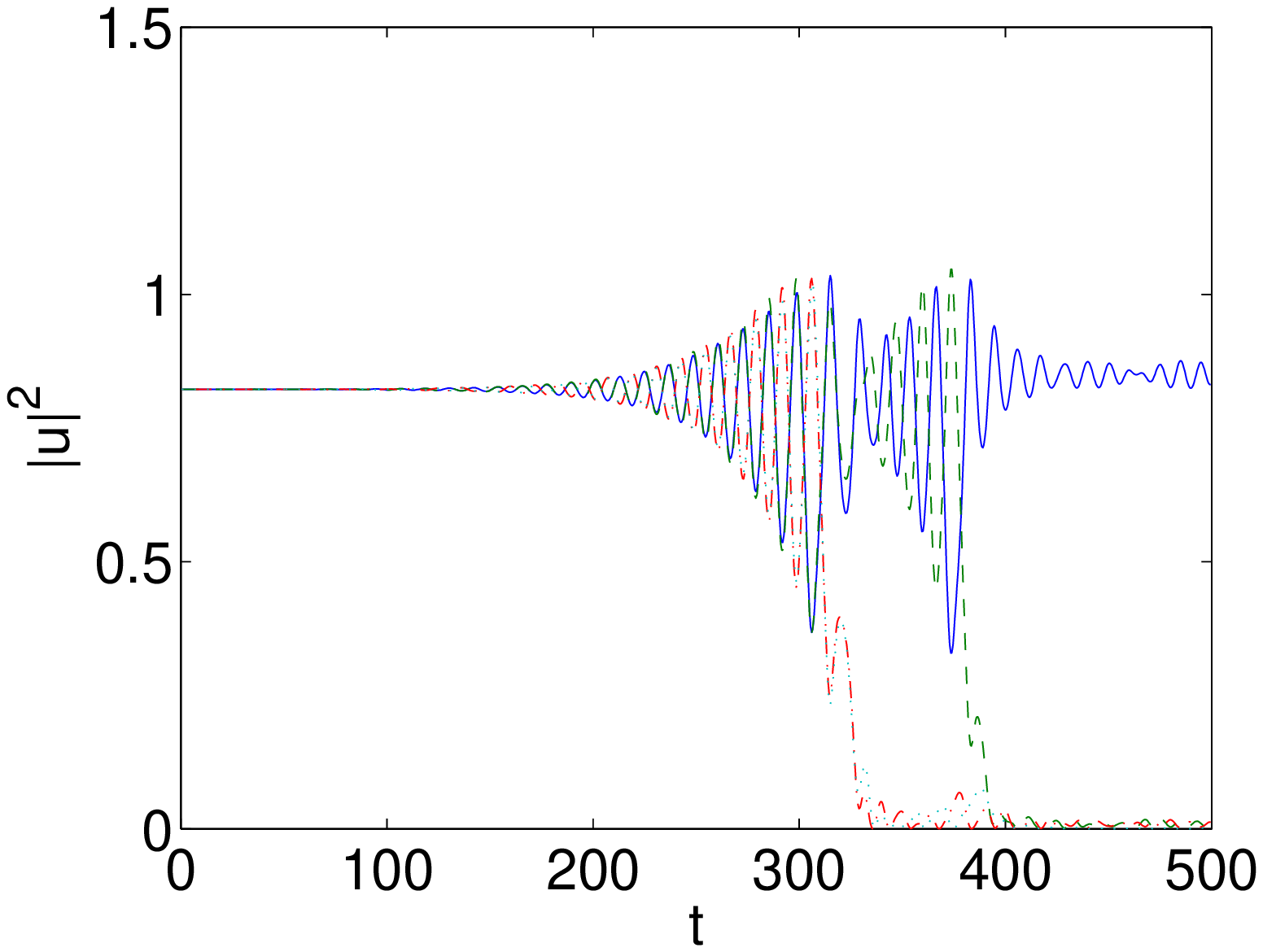}
\includegraphics[height=6cm,width=6cm]{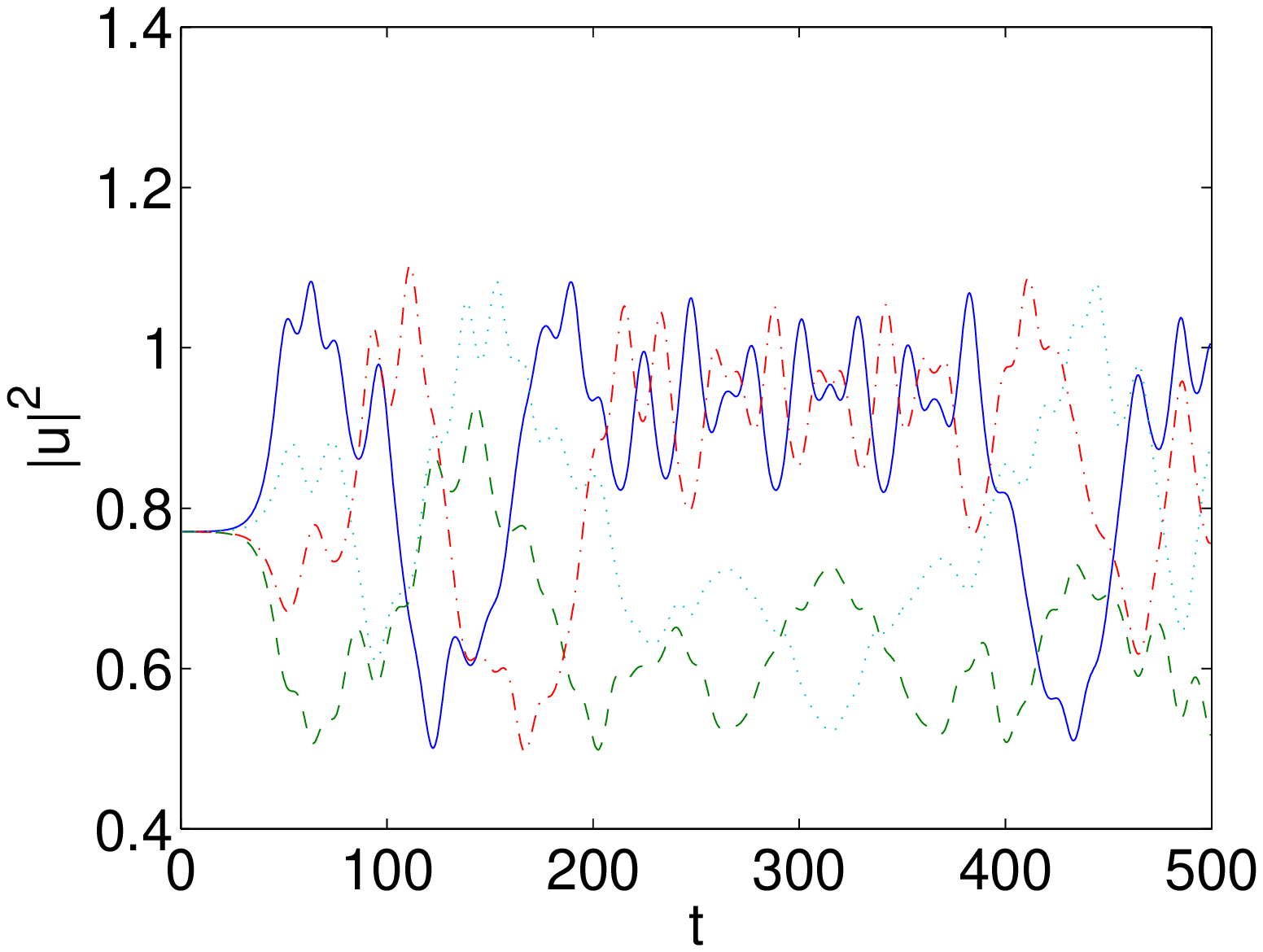} &
\end{tabular}
\end{center}
\par
\vskip-0.7cm 
\caption{(Color Online) Same as Fig. \ref{fig1a}, but now for the quadrupole
IS-IP mode with $C=0.08$ (left panels) and the quadrupole OS-IP mode 
with $C=0.05$ (right
panels). The left panels are for $t=200$, $t=300$ (top),
$t=400$ and $t=500$ (middle) and illustrate alongside the
bottom panel (containing the evolution of the principal 
four sites participating in the structure)
how the configuration eventually degenerates to a single-site
soliton. The right panels are for $t=50$, $t=150$, (top row) $t=250$
and $t=350$ (second row) and show together with the bottom panel the complex
oscillation (breathing) involved in the behavior of the 
quadrupole OS-IP mode for $C=0.05$.}
\label{fig2a}
\end{figure}

Figures \ref{fig2}-\ref{fig2a} show the quadrupolar mode
with four in-phase participating sites when centered
between lattice sites in the left panels of the figures.
This mode is theoretically predicted to have two 
imaginary (for small $C$) eigenvalue pairs with
\begin{eqnarray}
\lambda \approx 2 \sqrt{C} i
\label{QISIP}
\end{eqnarray}
and one imaginary pair with
\begin{eqnarray}
\lambda \approx \sqrt{8 C} i.
\label{QISIP2}
\end{eqnarray}
As a result, this mode (shown in the fourth row
panels of Fig. \ref{fig2} for $C=0.05$ and $C=0.1$)
is unstable due to the collision of the above eigenvalues
with the continuous spectrum occurring theoretically for
$C \approx 0.0477$, while in the numerical computations it happens
for $C \approx 0.047$. The outcome of the instability shown
in Fig. \ref{fig2a} for $C=0.08$ is the degeneration of
the quadrupolar mode into a single-site excitation.

\subsubsection{On-site, In-phase Mode}

The right panels of Figs. \ref{fig2}-\ref{fig2a} show the 
case of the on-site, in-phase mode. The latter is found to
always be unstable due to a real eigenvalue pair of
\begin{eqnarray}
\lambda \approx \pm 4 C
\label{QOSIP}
\end{eqnarray}
and a double, real eigenvalue pair of
\begin{eqnarray}
\lambda \pm \sqrt{12} C.
\label{QOSIP2}
\end{eqnarray}
This can also be clearly observed in the fourth and fifth
panels of Fig. \ref{fig2}, showing the mode and its stability for $C=0.05$.
The dynamical evolution of the unstable mode for $C=0.05$
is shown in the panels of Fig. \ref{fig2a}.
Both from the contour plots at the different times and from
the dynamical evolution of the main sites participating in
the structure, it can be inferred that the mode embarks in
an oscillatory breathing, without being ultimately destroyed
in this case.

\subsubsection{Inter-site, Out-of-Phase Mode}

\begin{figure}[tbp]
\begin{center}
\hskip-0.15cm
\begin{tabular}{cc}
\includegraphics[height=6cm,width=6cm]{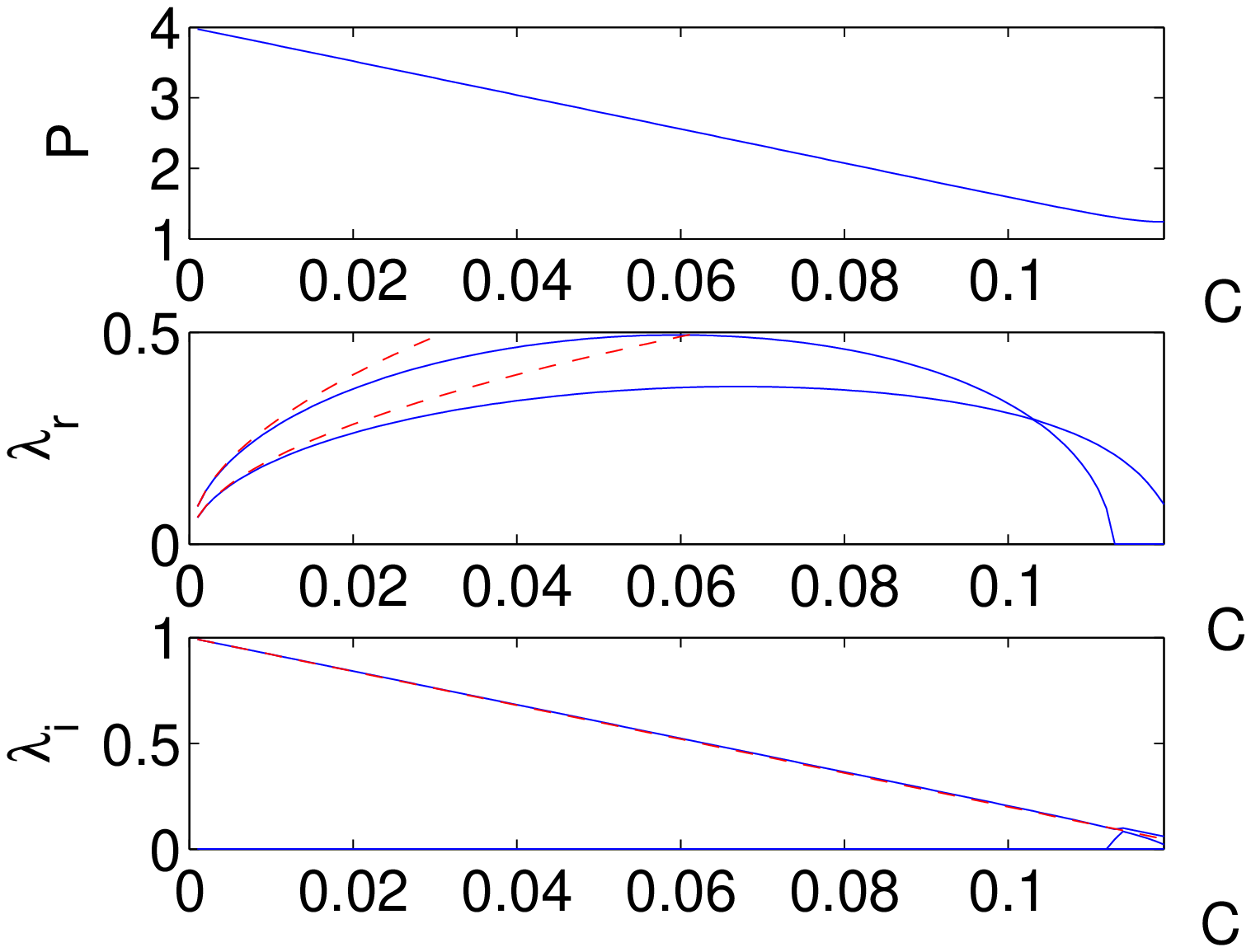}
\includegraphics[height=6cm,width=6cm]{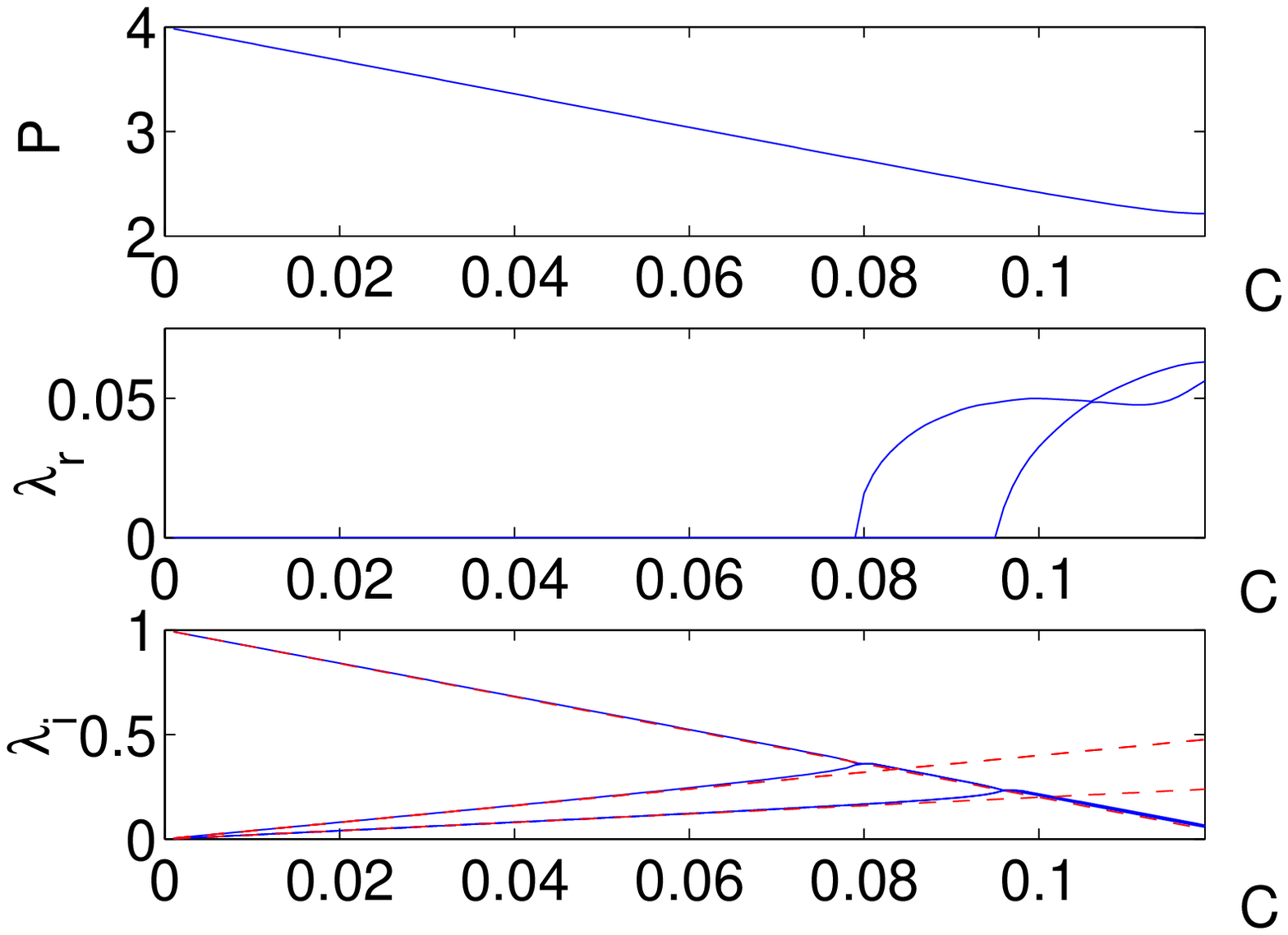} &  \\
\includegraphics[height=6cm,width=6cm]{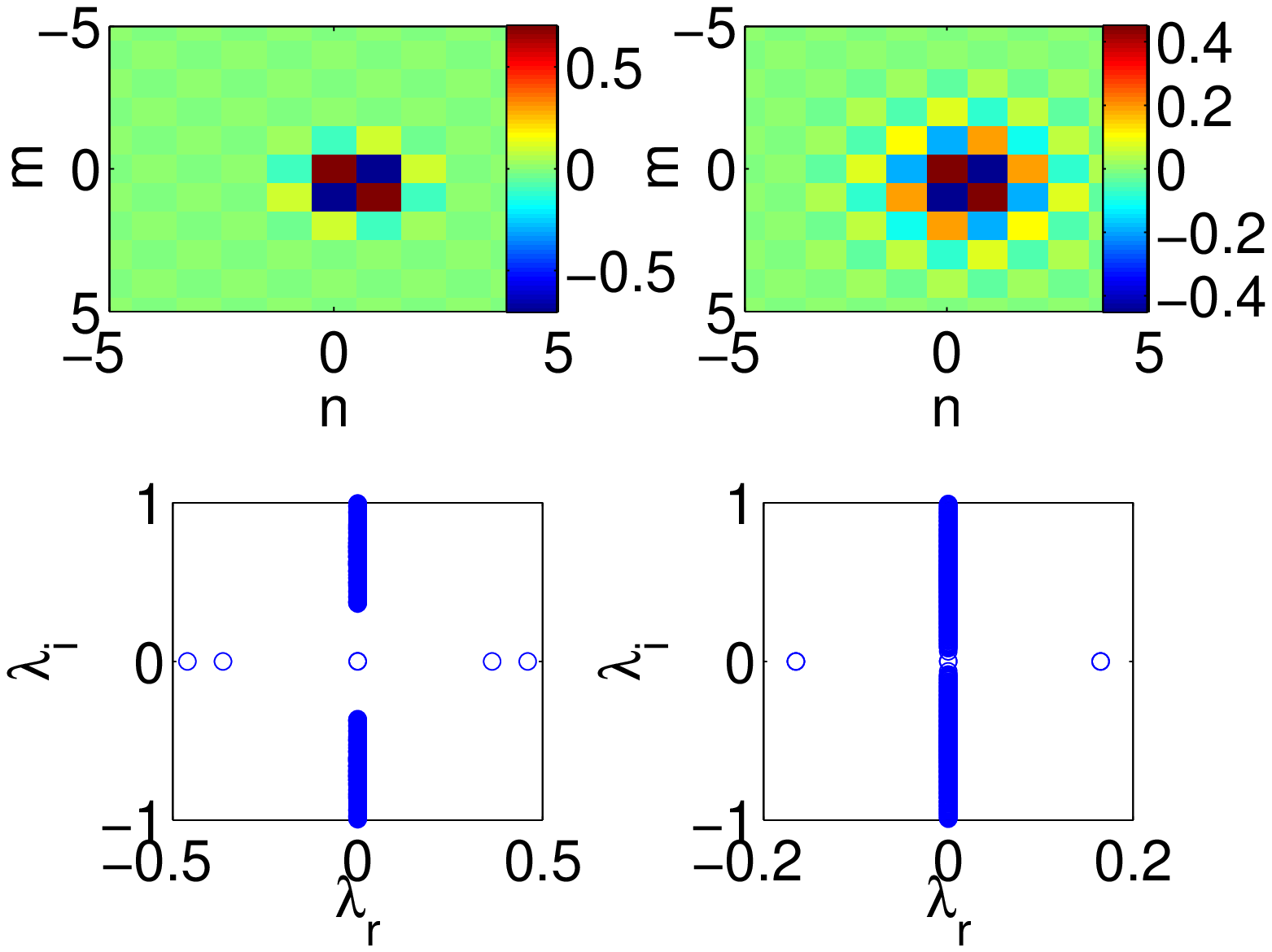}
\includegraphics[height=6cm,width=6cm]{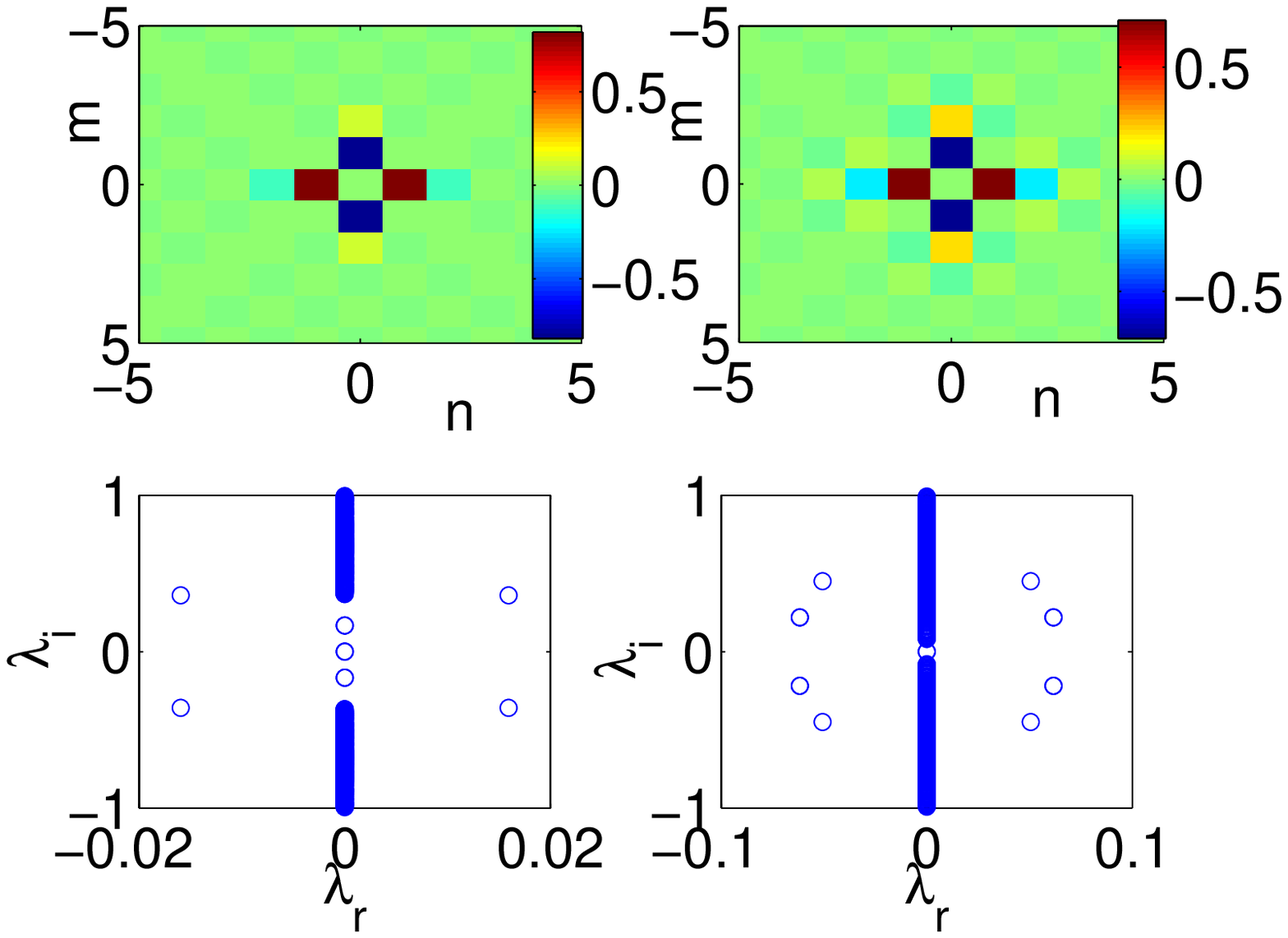} &  \\
\end{tabular}
\end{center}
\par
\vskip-0.7cm 
\caption{(Color Online) Similar to Fig. \ref{fig1} but for the
quadrupole IS-OP mode (left panels) and the quadrupole OS-OP mode
(right panels). The fourth and fifth rows show the modes and their
stability for $C=0.08$ and $C=0.116$ in each case.}
\label{fig2b}
\end{figure}

\begin{figure}[tbp]
\begin{center}
\hskip-0.15cm
\begin{tabular}{cc}
\includegraphics[height=6cm,width=6cm]{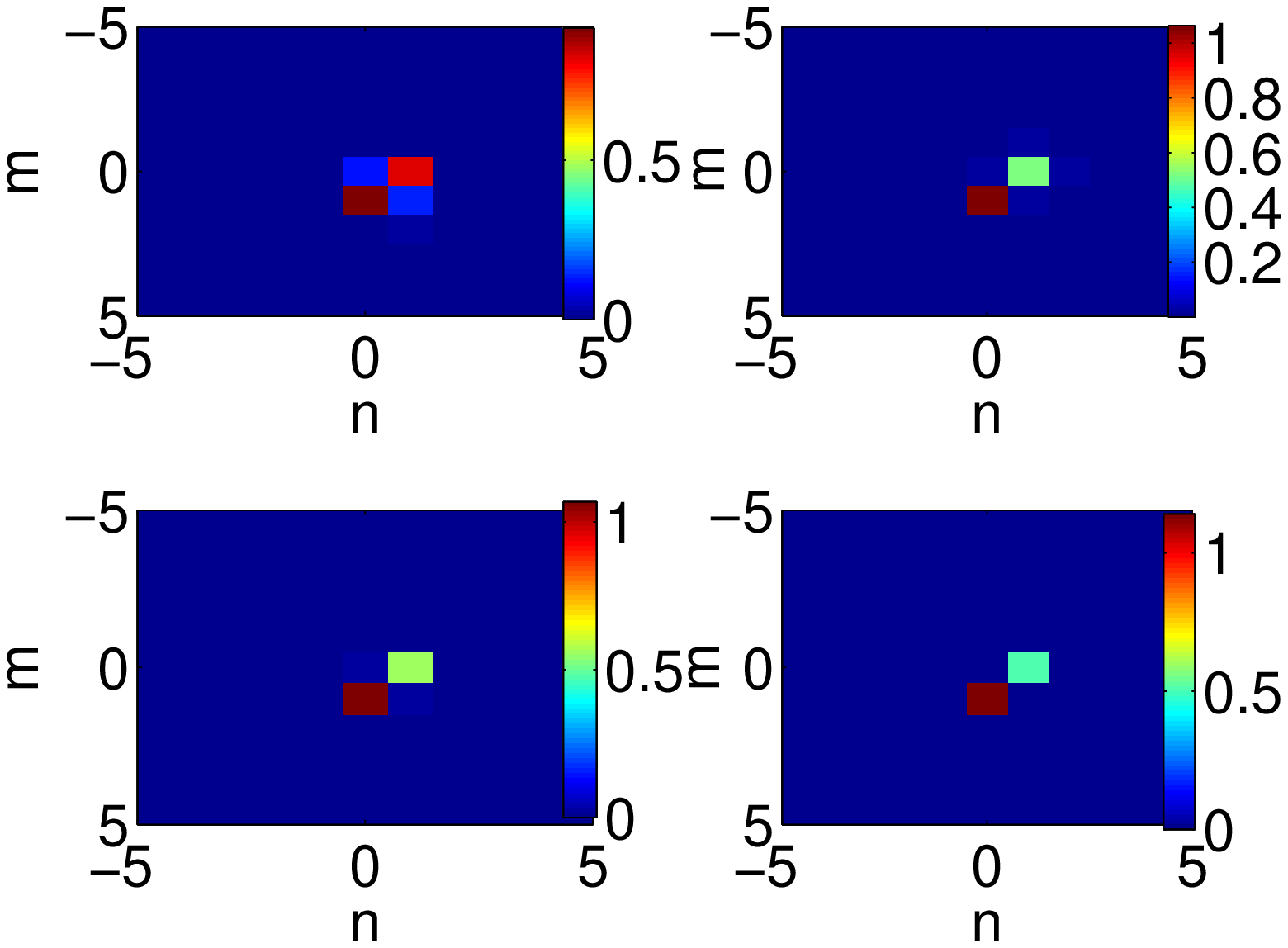}
\includegraphics[height=6cm,width=6cm]{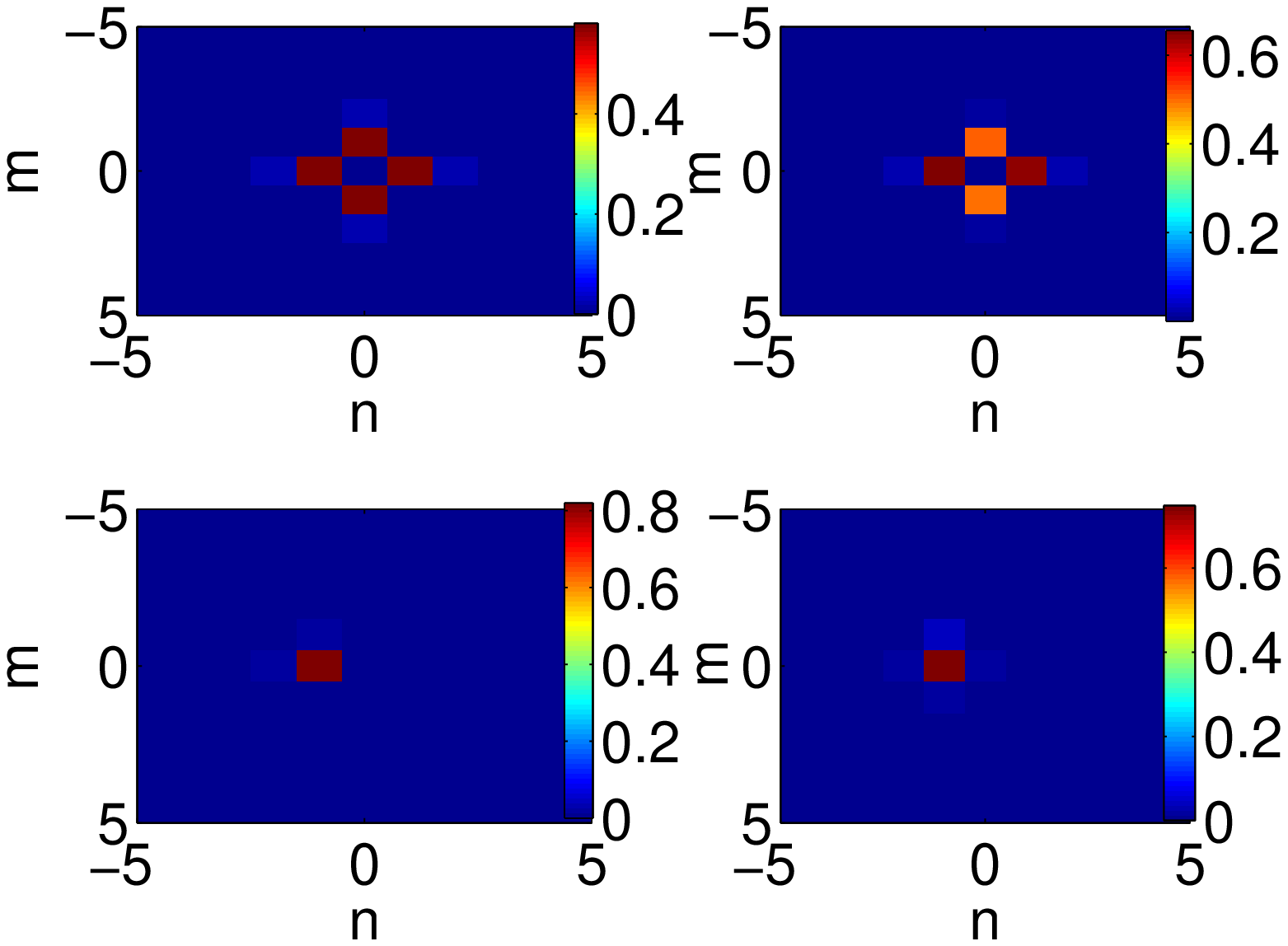} &  \\
\includegraphics[height=6cm,width=6cm]{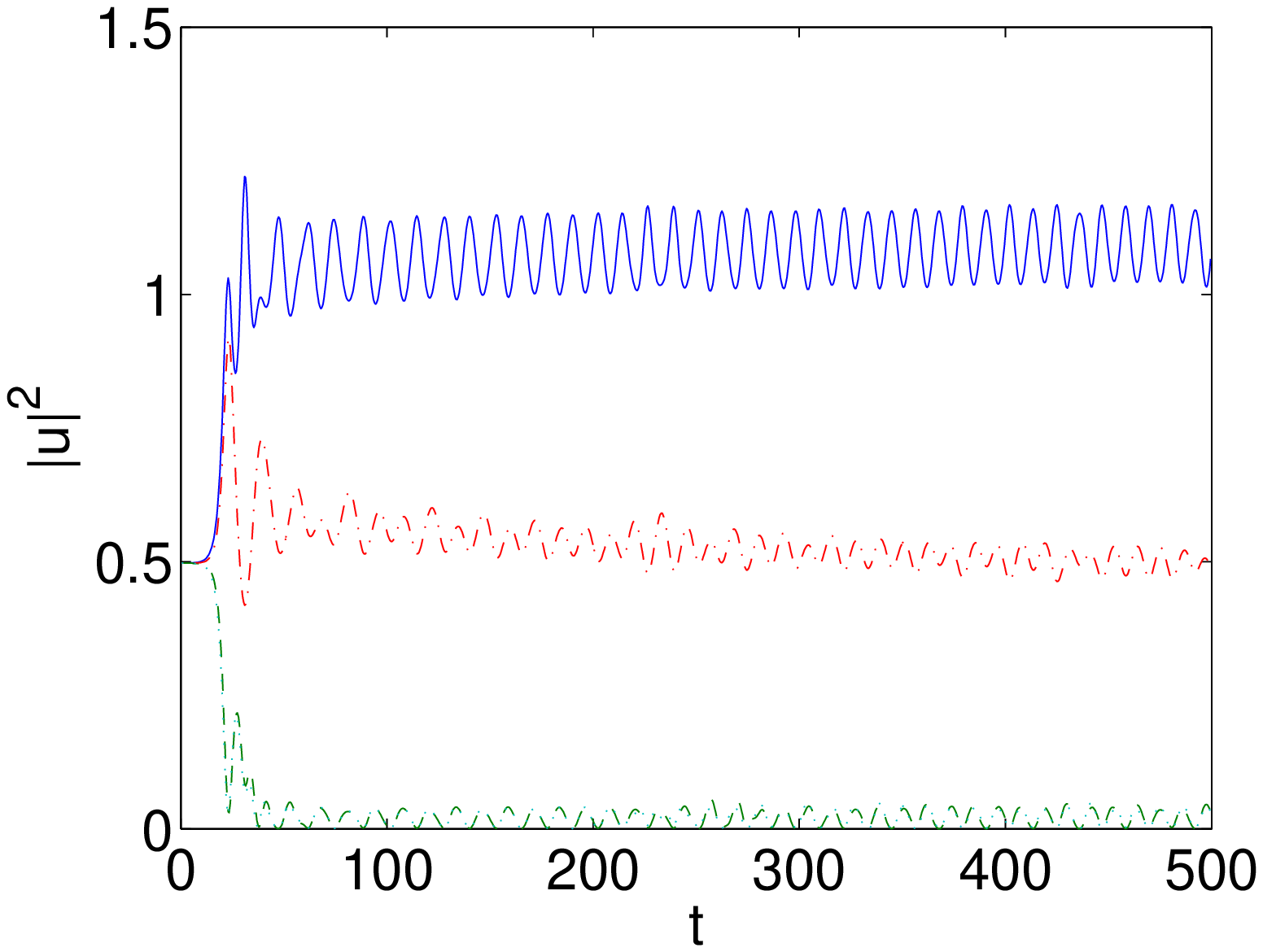}
\includegraphics[height=6cm,width=6cm]{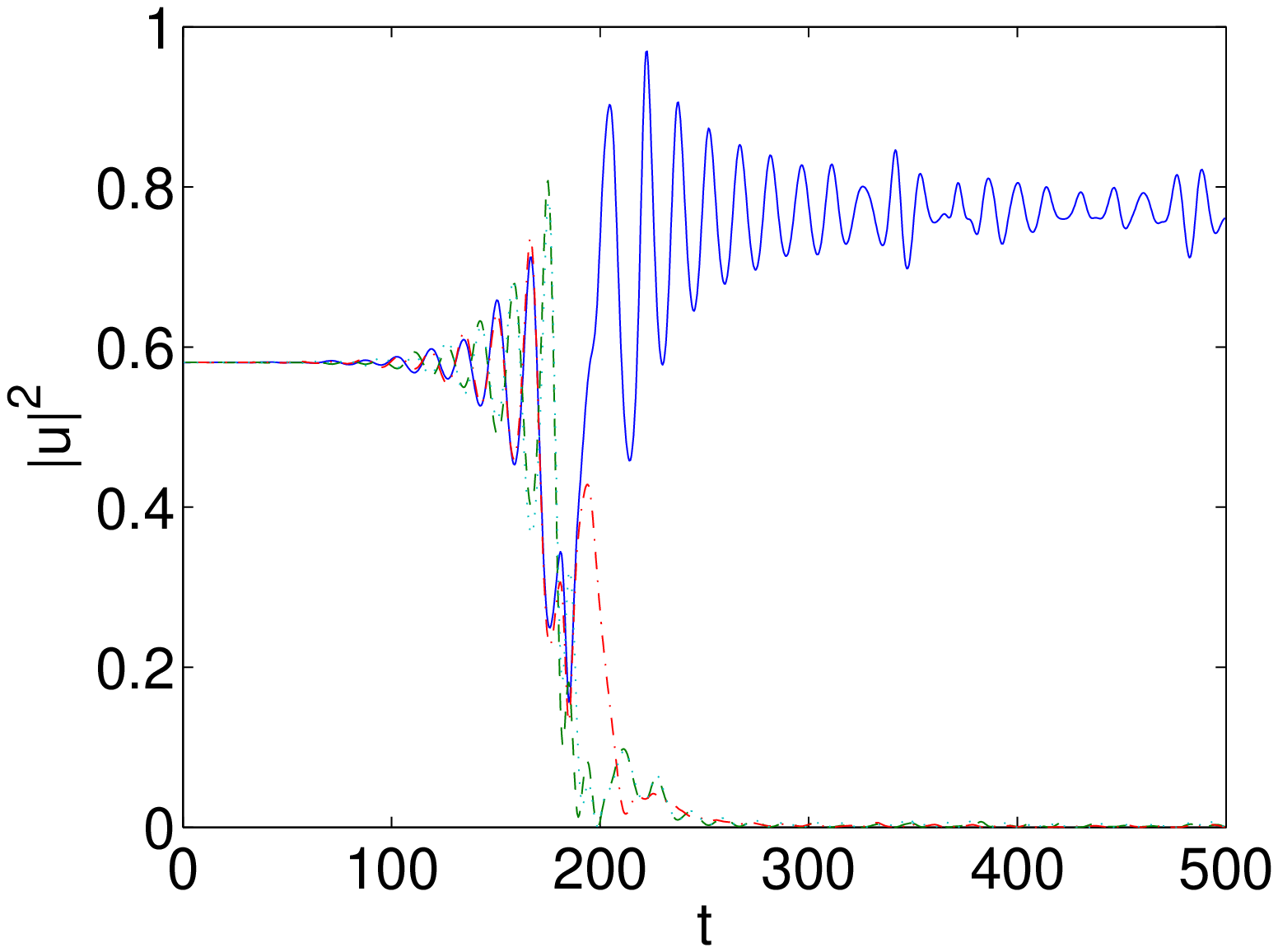} &
\end{tabular}
\end{center}
\par
\vskip-0.7cm 
\caption{(Color Online) Same as Fig. \ref{fig1a}, but for 
the evolution of the quadrupole IS-OP mode with $C=0.08$ (left panels)
the quadrupole OS-OP mode with $C=0.1$ (right panels). The former
is shown for $t=25$, $t=50$ (top row), $t=150$ and $t=250$ (second row)
substantiating
(together with the bottom panel showing the principal four sites
of the branch) its resulting into a long-lived, breathing two-site waveform.
The latter is shown for $t=50$, $t=150$ (top row), $t=250$ and $t=350$
(second row)
indicating its degeneration into a single-site configuration.}
\label{fig2c}
\end{figure}

We next consider the case of the IS-OP mode in Figs. \ref{fig2b}-\ref{fig2c}.
Our analytical results for this mode show that for small values of
$C$, we should expect to find it to be immediately unstable due to
three real pairs of eigenvalues, namely a single one with
\begin{eqnarray}
\lambda \approx \pm \sqrt{8 C}
\label{QISOP1}
\end{eqnarray}
and a double one with
\begin{eqnarray}
\lambda \approx \pm 2
\sqrt{C}.
\label{QISOP2}
\end{eqnarray}
This expectation is once again confirmed by the numerical results
of the left panel of Fig. \ref{fig2b}. The fourth and fifth panels
show the mode and the spectral plane of its linearization 
for the cases of $C=0.08$ and $C=0.116$. The dynamical evolution of this
mode also gives an interesting result, in that it produces, upon
manifestation of the instability, a long-lived, two-site oscillatory
mode, as is illustrated in the left panels of Fig. \ref{fig2c}
for $C=0.08$.

\subsubsection{On-site, Out-of-phase Mode}

Finally, the last one among the quadrupolar modes is the 
OS-OP mode, examined in the right panels of Figs. \ref{fig2b}-\ref{fig2c}.
Our theoretical analysis predicts that this mode should have
a double imaginary eigenvalue pair of
\begin{eqnarray}
\lambda\approx \pm 2 C i
\label{QOSOP1}
\end{eqnarray}
and a single imaginary pair of
\begin{eqnarray}
\lambda \approx 4 C i.
\label{QOSOP2}
\end{eqnarray}
These, in turn, imply that the mode is stable for small $C$, but
becomes destabilized upon collision of the larger one among these
eigenvalues with the continuous band of phonons. This is numerically
found to occur for $C \approx 0.08$, while it is theoretically
predicted, based on the above eigenvalue estimates, to take place
for $C =0.083$. The mode's stability analysis is
shown in the fourth and fifth panel
of Fig. \ref{fig2b} for $C=0.08$ and $C=0.116$; for $C=0.1$, and
its dynamical evolution is examined in the right panels of Fig.
\ref{fig2c}. In this case, we do find that the mode essentially
degenerates to a single-site solitary wave.

\subsection{Vortex Configuration}

\begin{figure}[tbp]
\begin{center}
\hskip-0.15cm
\begin{tabular}{cc}
\includegraphics[height=6cm,width=6cm]{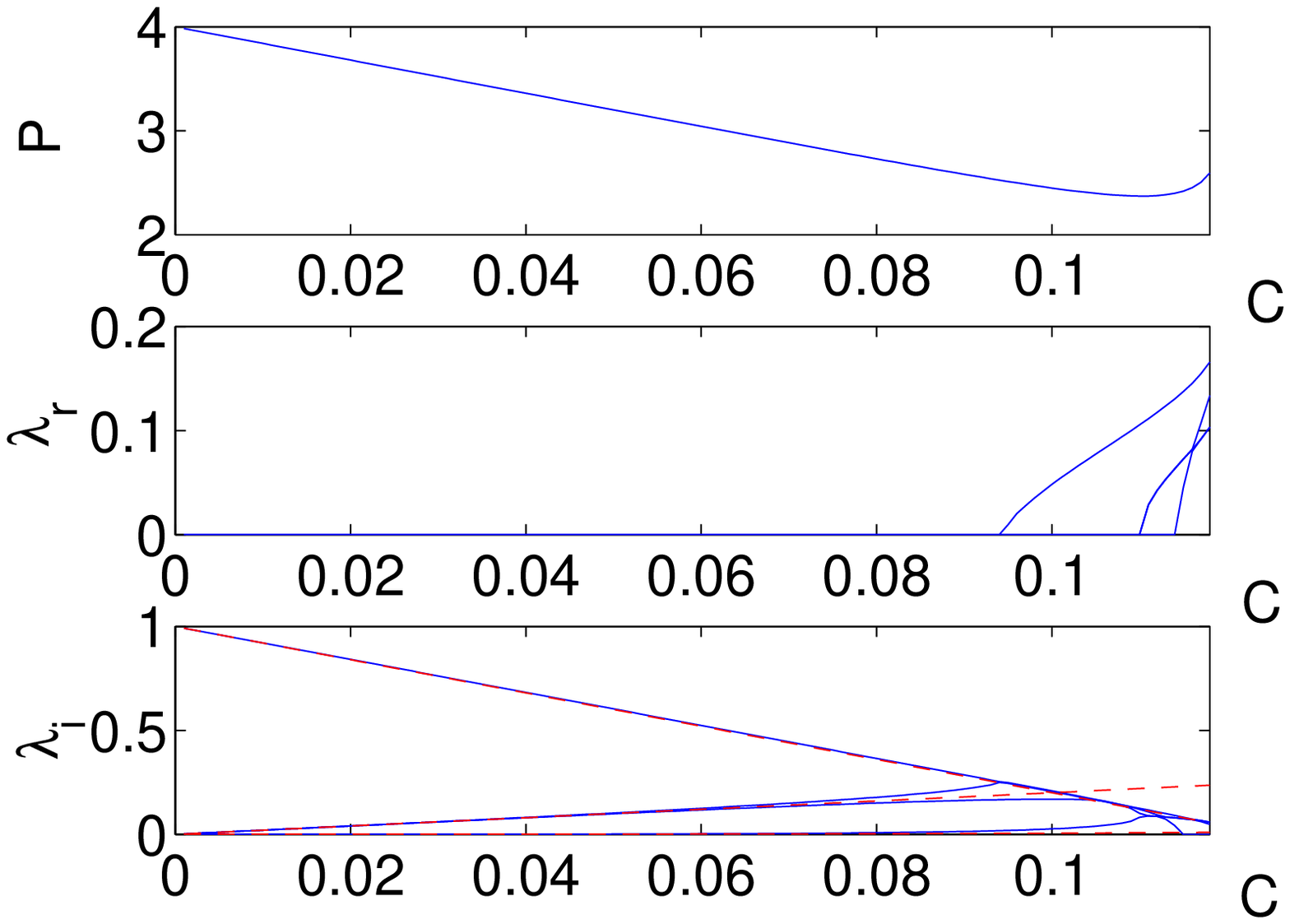}
\includegraphics[height=6cm,width=6cm]{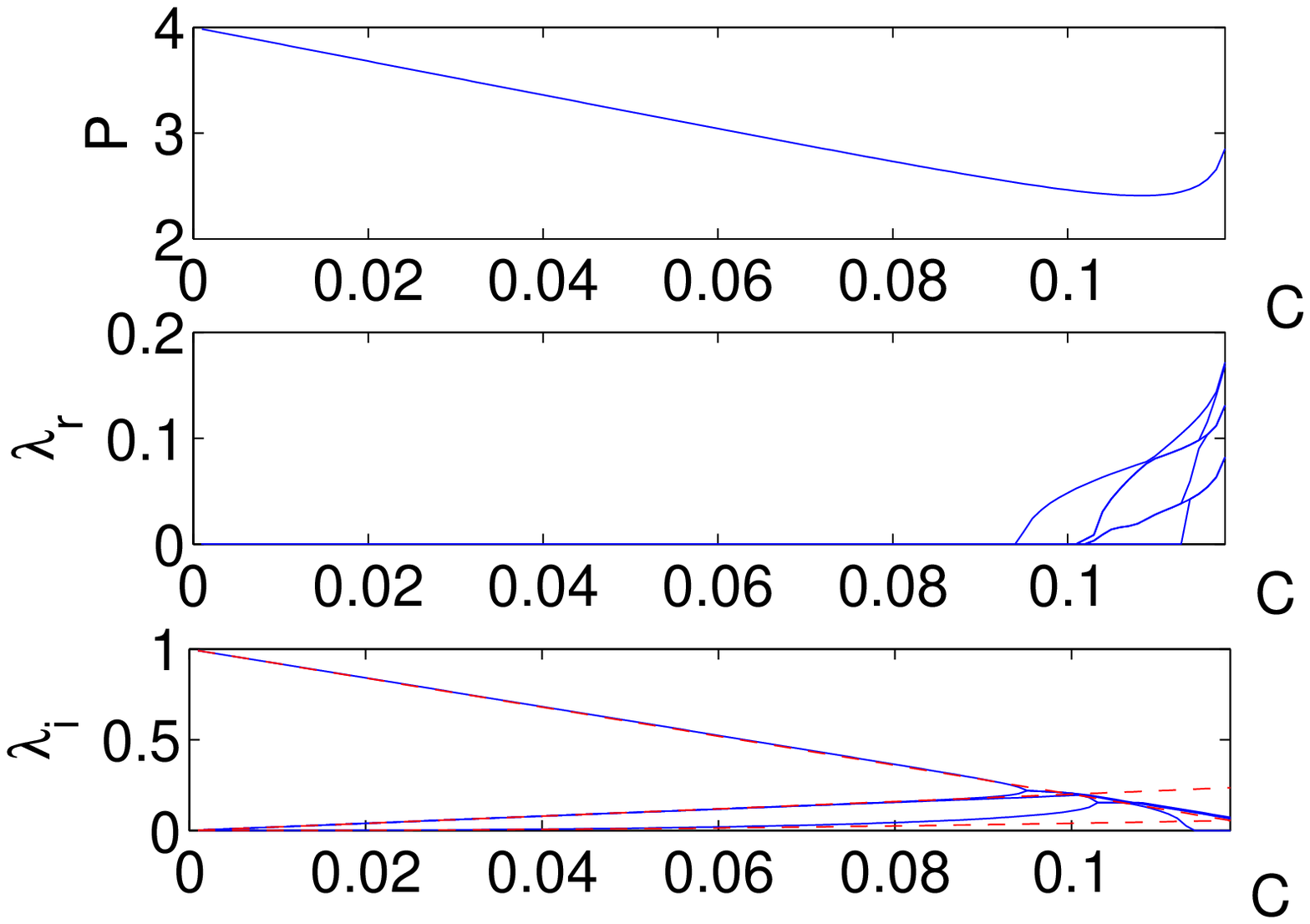} &  \\
\includegraphics[height=6cm,width=6cm]{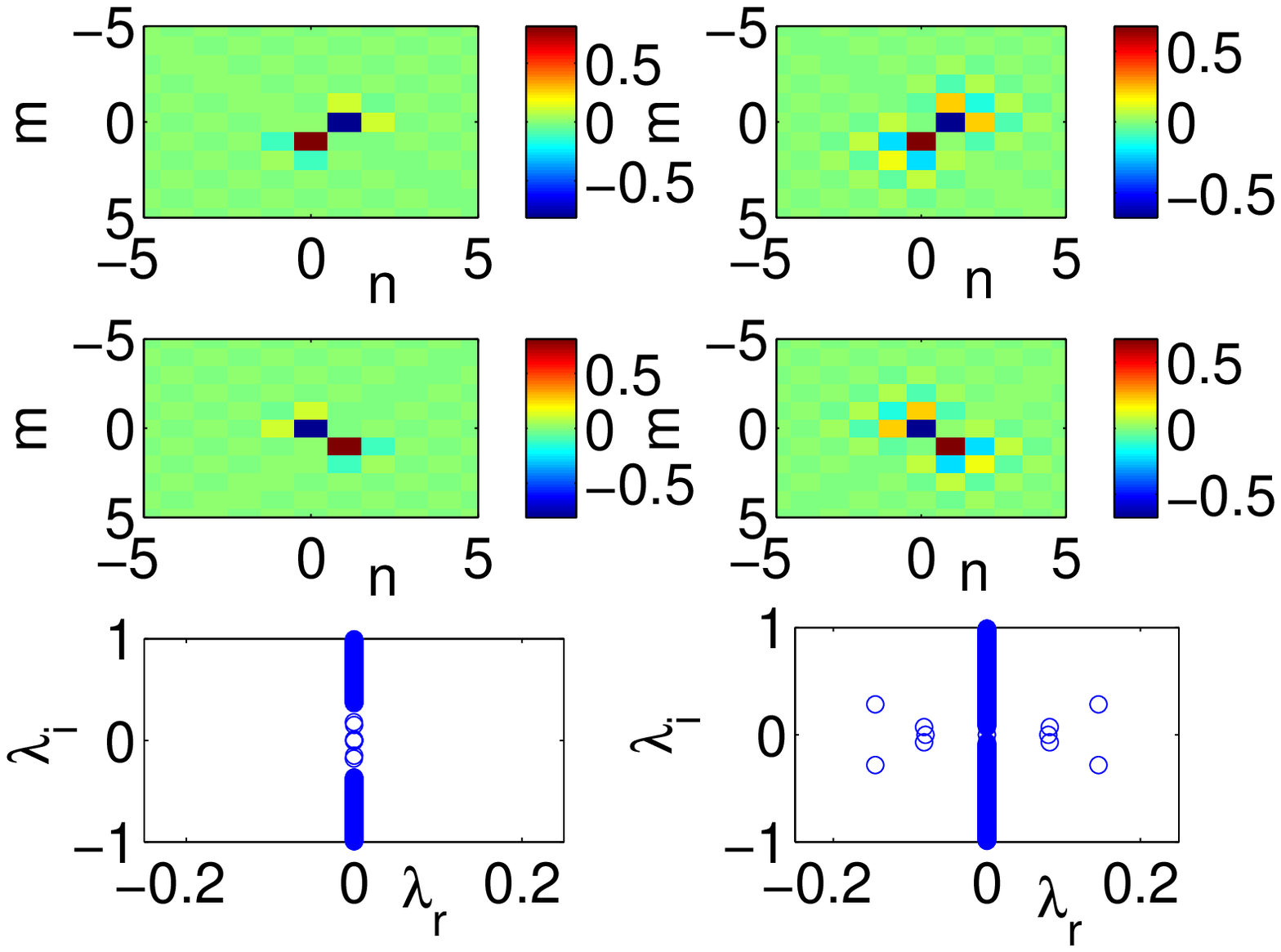}
\includegraphics[height=6cm,width=6cm]{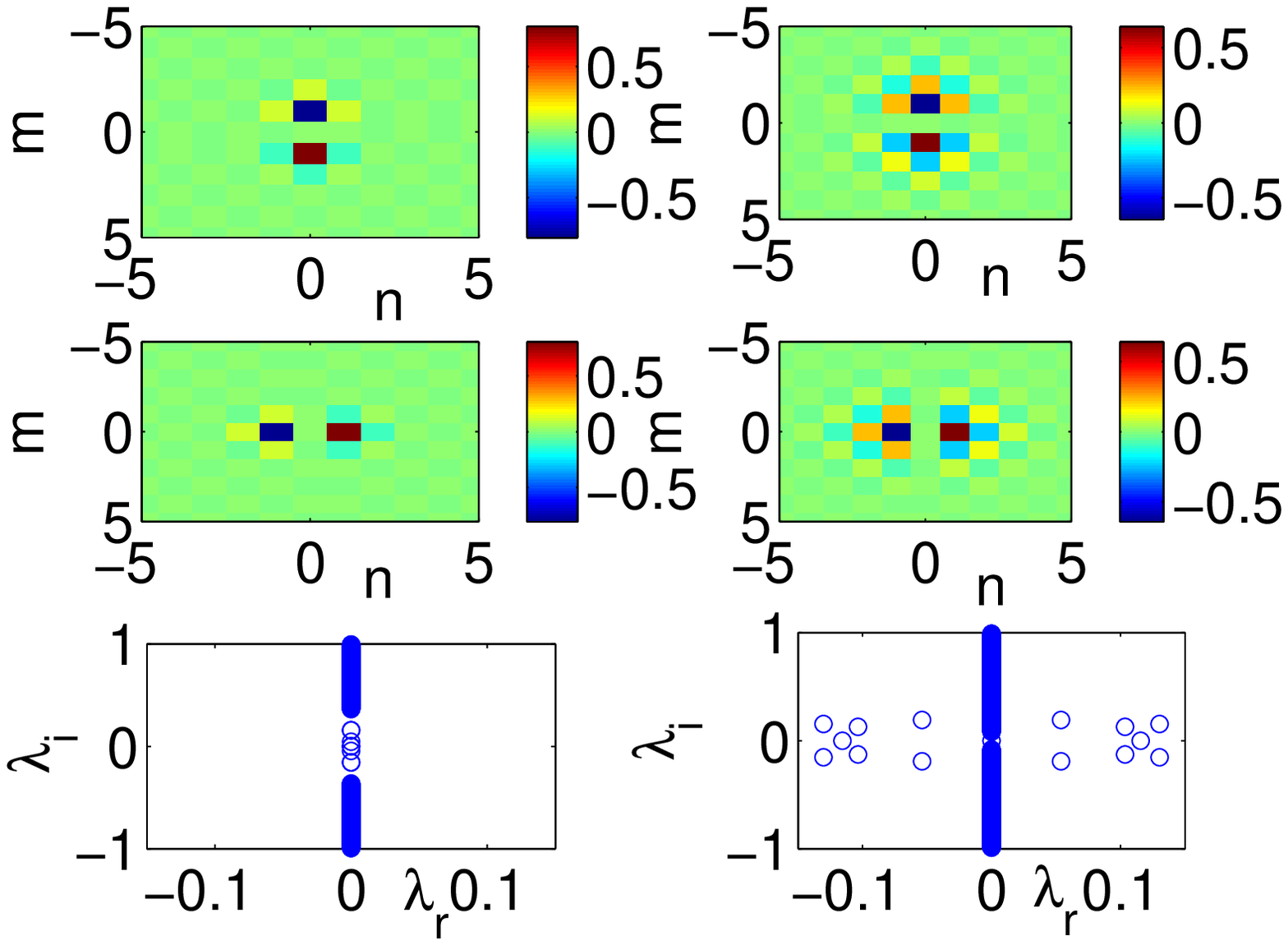} &
\end{tabular}\end{center}
\par
\vskip-0.7cm 
\caption{(Color Online) The same features as in Figure 1
are shown here for the IS vortex of topological charge $S=1$
(left) and the OS vortex of $S=1$ (right). In this case both
the real (fourth line) and imaginary (fifth line) parts of the
solution are shown (and their stability in the sixth line) for
$C=0.08$ and $C=0.116$.}
\label{fig3}
\end{figure}

\begin{figure}[tbp]
\begin{center}
\hskip-0.15cm
\begin{tabular}{cc}
\includegraphics[height=6cm,width=6cm]{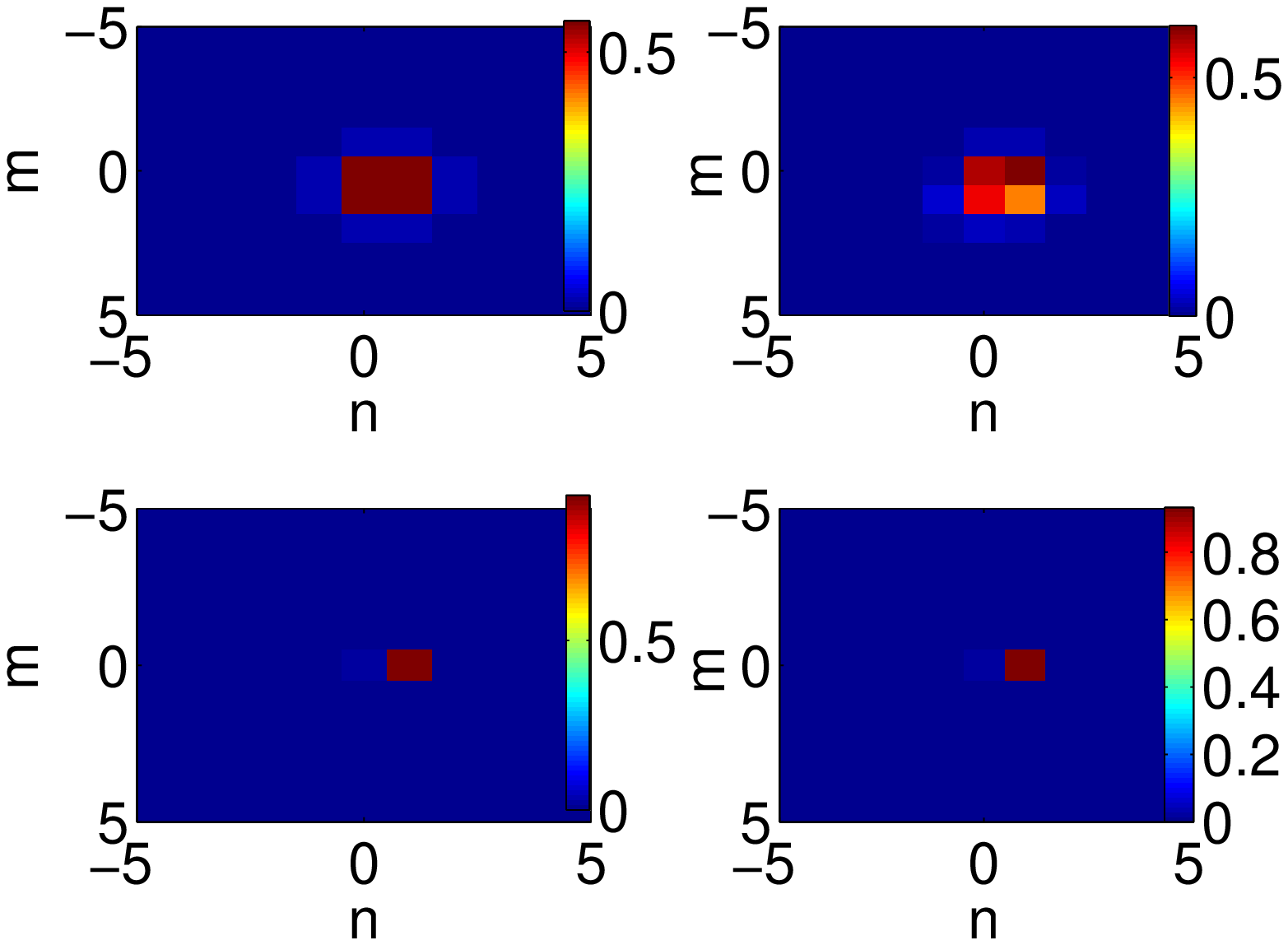}
\includegraphics[height=6cm,width=6cm]{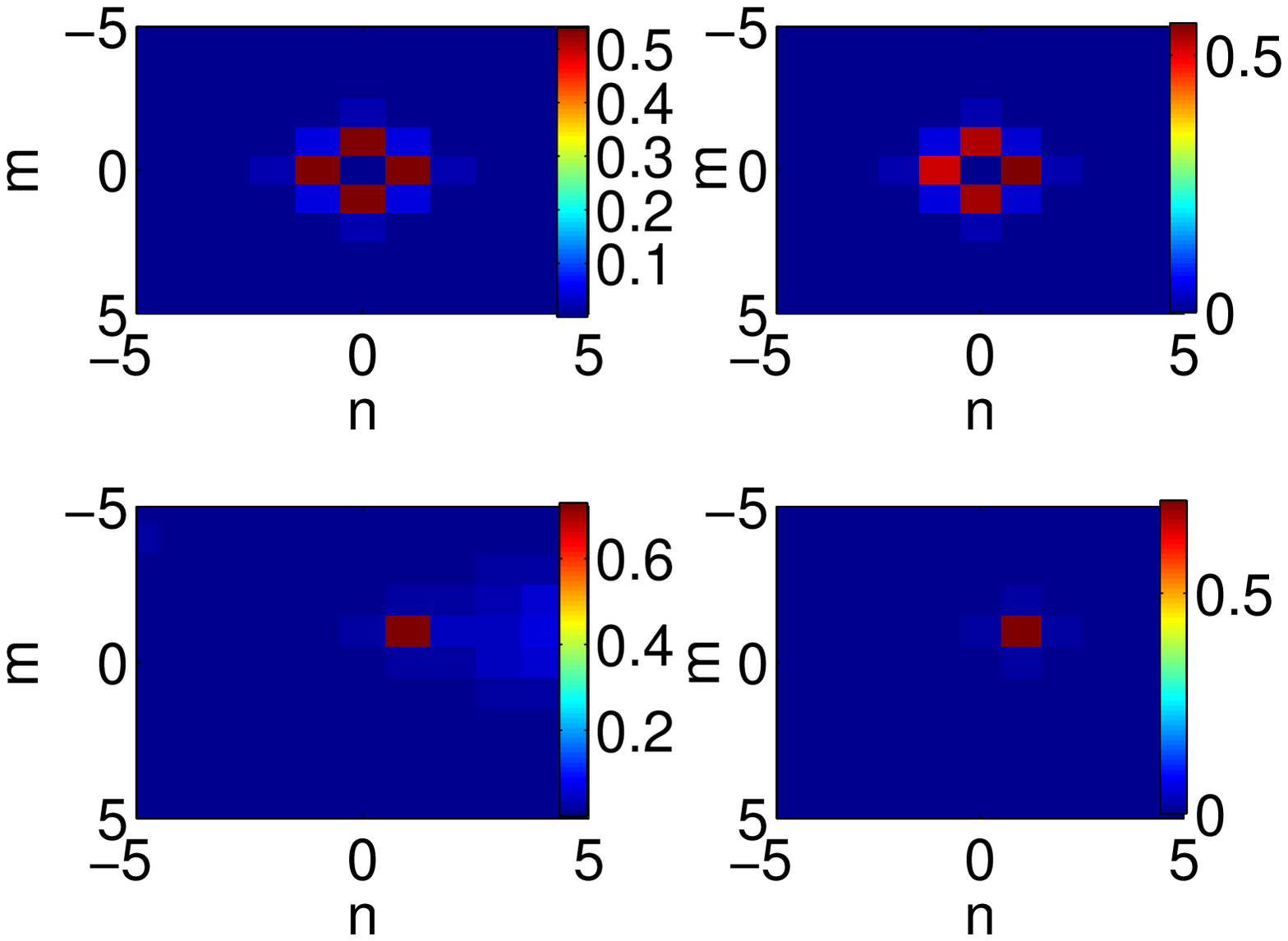} &  \\
\includegraphics[height=6cm,width=6cm]{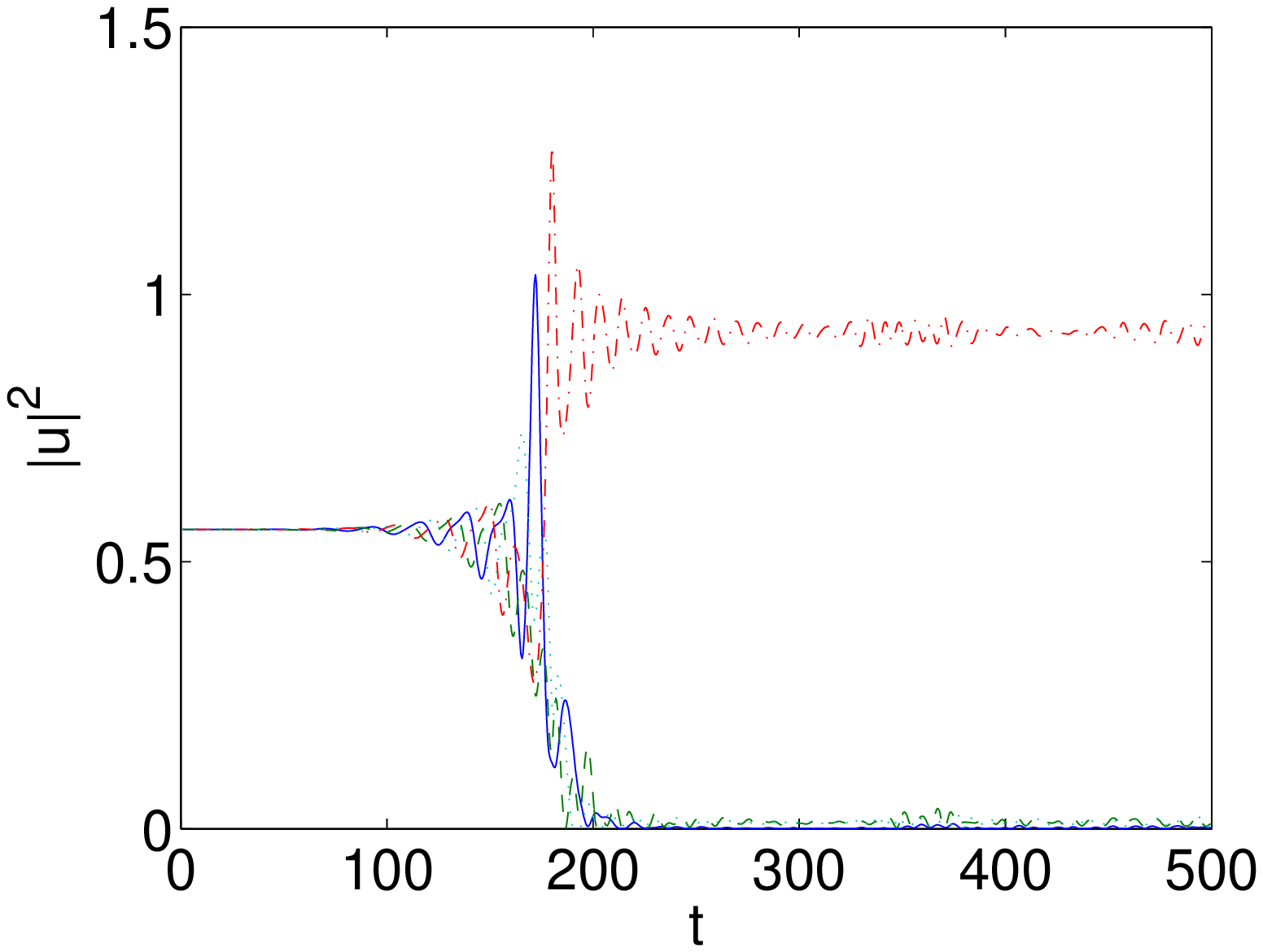}
\includegraphics[height=6cm,width=6cm]{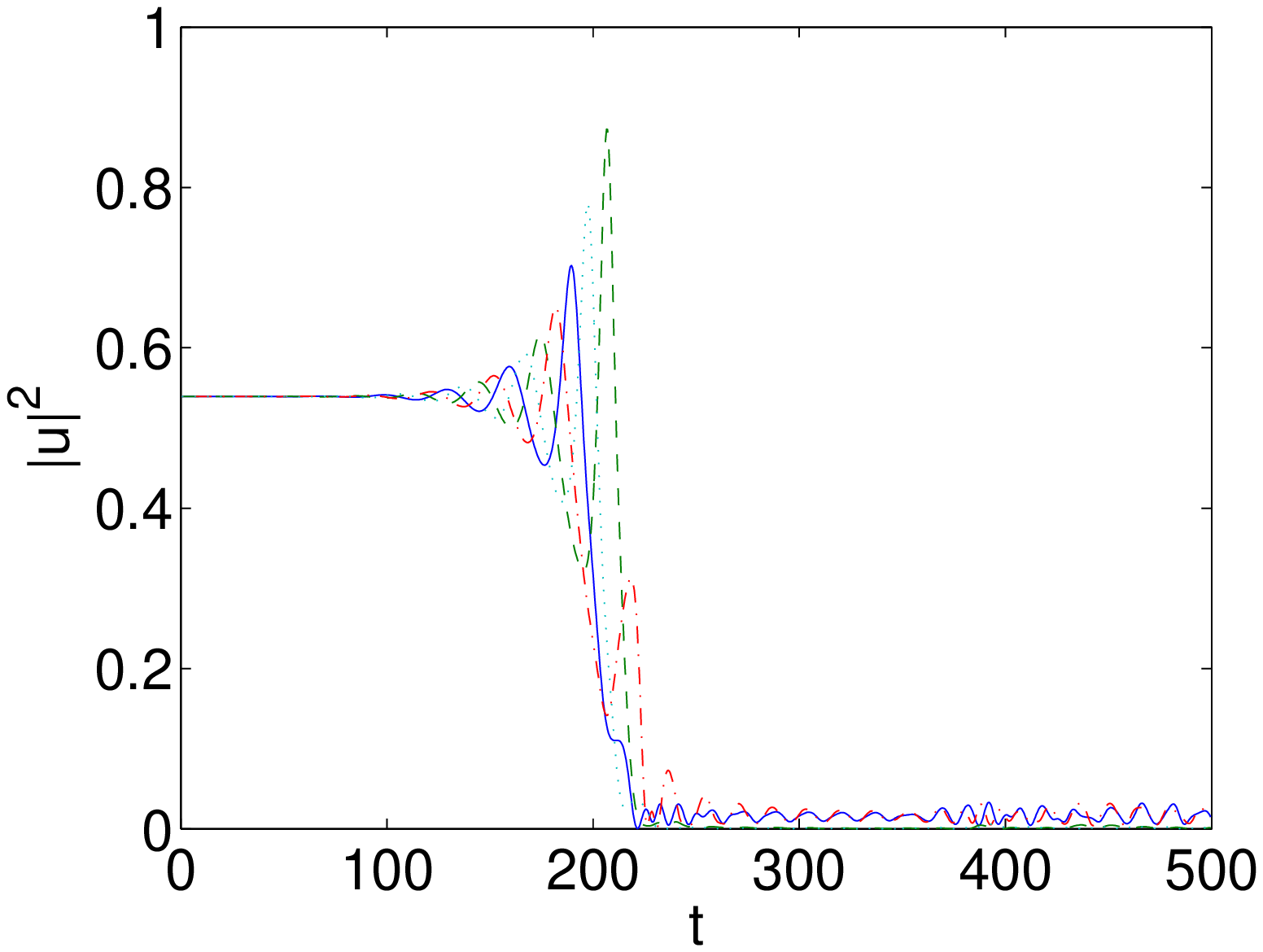} &
\end{tabular}
\end{center}
\par
\vskip-0.7cm 
\caption{(Color Online) Same as in Fig. \ref{fig1a} but
for the dynamical evolution of the IS (left panels) and
the OS (right panels) vortex of topological charge
$S=1$. Both cases are shown for $C=0.1$. Both the left
and right panels of evolution are for $t=50$, $t=150$ (top row),
$t=250$ and $t=350$ (second row) and alongside the bottom panels showing
the principal four sites of the vortex they show how
the vortices transform themselves into fundamental
solitary waves centered on a single site (in the latter case,
interestingly, the site of the largest excitation of the ensuing
wave is not one among the four principal excitation sites of
the original OS vortex).}
\label{fig3a}
\end{figure}

\subsubsection{Inter-site Vortices}

Finally, Figs. \ref{fig3}-\ref{fig3a} show similar features,
but now for the IS (left panels) and OS (right panels) vortex 
solutions \cite{dnls,dep}.
The former has a theoretically predicted double pair of eigenvalues
\begin{eqnarray}
\lambda \approx \pm 2 C i.
\label{VIS}
\end{eqnarray}
leading to
an instability upon collision with the continuum band for 
$C\geq 0.095$ ($C\geq 0.1$ theoretically). In this
case, there is also an eigenvalue of higher
order 
\begin{eqnarray}
\lambda \approx \pm 4 C^2 i. 
\label{VIS2}
\end{eqnarray}
which obviously depends more weakly on $C$.
The fourth and fifth panels of Fig. \ref{fig3} show the
real and the imaginary parts of the vortex configuration
for $C=0.08$ and $C=0.116$ and the sixth panel the
corresponding spectral planes. The dynamical evolution
of the vortex of topological charge $S=1$, for $C=0.1$ is shown
in the left panels of Fig. \ref{fig3a}, indicating that the
vortex also, upon the occurrence of the oscillatory instability,
becomes a single-site solitary wave.

\subsubsection{On-site Vortices}

On the other hand, the OS vortices are shown in the right panels
of Figs. \ref{fig3}-\ref{fig3a}. In this case, we theoretically
find that the vortex, for small $C$, should have a double pair
of eigenvalues
\begin{eqnarray}
\lambda \approx 2 C i
\label{VOS}
\end{eqnarray}
and a single, higher order pair of eigenvalues
\begin{eqnarray}
\lambda \approx \pm \sqrt{32} C^3 i.
\label{VOS2}
\end{eqnarray}
The former eigenvalue pairs, upon collision with the continuous
spectrum, lead to an instability, theoretically predicted to 
occur at $C=0.1$ and numerically found to happen for $C \approx 0.095$.
The on-site mode (and its stability) is shown in the fourth-sixth right
panels of Fig. \ref{fig3} for $C=0.08$ and $C=0.116$. Its evolution
(for $C=0.1$) is shown in the right panel of Fig. \ref{fig3a}, where
it is again seen that the mode degenerates from an $S=1$ to an $S=0$
structure, i.e., a single-site solitary wave with no topological charge.

\subsection{General Principles Derived From Stability Considerations}

It is interesting to note as an over-arching conclusion that the
stability intervals of the defocusing structures are different from
those of their focusing counterparts (especially when they are stable
close to the AC limit) because of the collisions with the continuous
spectrum
band edge; the latter is at $\lambda=\Lambda$ in the focusing case, while it
is at $\lambda=\Lambda-8C$ in the defocusing setting. Another similarly general
note is an immediate inference on whether the structures are stable or
not; this can be made based on the knowledge of whether their focusing 
counterparts are stable or not and the transformation from the former
to the latter through the staggering transform: $u_{n,m}=
(-1)^{n+m} v_{n,m}$. For instance, IP two-site configurations (both OS and
IS) are known to be generically unstable in the focusing regime \cite{dep}; 
through the staggering transformation,
OS-IP of the focusing case 
remains OS-IP in the defocusing, while IS-IP of the focusing 
becomes IS-OP in the defocusing.
Hence, these two should be expected to be always 
unstable, while the remaining two (OS-OP in both focusing and
defocusing and IS-OP of the focusing, which becomes IS-IP in the
defocusing) should similarly be expected
to be linearly stable close to the AC-limit, as is indeed observed. 
Notice that, interestingly enough, for the vortex states the staggering
transformation indicates that the stability is not modified between
the focusing and defocusing cases. This is because for an IS vortex, it
transforms an $S=1$ state into an $S=-1$ state (which is equivalent
to the former, in terms of stability properties), while the OS vortex
remains unchanged by the transformation.
However, as mentioned above, these considerations are not sufficient to
compute the instability thresholds for initially stable modes, among other
things. They do, nonetheless, provide a guiding principle for inferring
the near-AC limit stability of the defocusing staggered states, based
on their focusing counterparts.

\section{Conclusions and Future Challenges}

In this paper, we have studied in detail some of the principal
multi-site solitary wave structures that emerge in the context
of defocusing nonlinearities, examining, in particular,  
dipole, quadrupole and vortex configurations. We have illustrated
which ones among these states can potentially be stable 
(e.g. IS-IP and OS-OP for both dipoles and quadrupoles, as
well as the vortices) and those that will {\it always} be unstable
(e.g. IS-OP and OS-IP modes for both dipoles and quadrupoles).
We have also provided detailed analytical estimates of the
stability eigenvalues associated with these modes, in 
very good agreement with the observed numerical results.
The analytical calculations also empower us to identify,
even for the stable (close to the AC-limit) 
modes, the relevant intervals of stability
of those waveforms. We have corroborated our analytical 
calculations with detailed computations that identify the
corresponding modes and numerically analyze their linear
stability. In addition, for each of the modes, we have shown some
typical examples of their dynamical evolution, when they become
unstable (either directly, or subsequently due to eigenvalue 
collisions). 

These results offer immediate suggestions for experiments
in arrays of optical waveguides, Bose-Einstein condensates
(e.g. of $^{87}$Rb or $^{23}$Na, which feature repulsive 
interactions amounting, at the mean-field level,
to a defocusing nonlinearity) mounted on a deep optical
lattice. In the latter case, the nodes of the lattice
considered herein
would correspond to BEC droplets in the respective
wells of the optical potential. Finally, they are also
suggestive of similar experiments in the recently and
rapidly growing theme of photorefractive crystal lattices
(where, however, the nonlinearity is slightly different,
featuring a saturable form).

We close by suggesting that these results also indicate
that higher charge configurations \cite{dep,zhigang_pre}
may similarly be possible and could potentially also
be stable in a defocusing setting, similarly to the
$S=1$ states discussed above. It would certrainly be
of interest to examine such states in the near future,
as well as to study the effect of additional components \cite{dep1}
(i.e., multi-component states, relevant to the above
optical settings when multiple
polarizations are present, or to BECs when multiple hyperfine states
are studied), or that of higher-dimensional
structures \cite{ricardo}.

\vspace{5mm}
{\bf Acknowledgements.} PGK gratefully acknowledges the support
of NSF through the grants DMS-0204585,
DMS-CAREER, DMS-0505663 and DMS-0619492. ZC was supported by
AFOSR, NSF, PRF and NSFC.

\end{document}